\documentclass[journal]{IEEEtran}
\usepackage[keeplastbox]{flushend}
\usepackage{xcolor}
\usepackage{amsmath}
\usepackage{amsthm}
\theoremstyle{plain}

\usepackage{cite}
\usepackage{amsfonts,amssymb}
\usepackage{graphicx}
\usepackage{subfigure}
\usepackage{multirow}
\usepackage{enumerate}
\usepackage[shortlabels]{enumitem}
\usepackage[T1]{fontenc}% optional T1 font encoding
\usepackage{stfloats}
\usepackage{booktabs}
\usepackage[noend]{algorithmic}
\usepackage{tikz}
\usepackage[subnum]{cases}
\usepackage{algorithm}
\usepackage{array}
\usepackage{color}
\usepackage{colortbl}
\usepackage{mathtools}
\usepackage{url}
\usepackage{float}
\usepackage{stfloats}

\usetikzlibrary{calc,trees,positioning,arrows,arrows.meta,arrows.spaced,chains,shapes.geometric,shapes.misc,%
    decorations.pathreplacing,positioning,decorations.pathmorphing,shapes,%
    matrix,shapes.symbols}
    
\tikzset{
    decision/.style={shape aspect=2, diamond, draw, minimum size=1.2cm, inner sep=0pt, align=center},
    text/.style={ diamond, draw, minimum size=1.2cm, inner sep=0pt, align=center},
    line/.style={draw, -{latex}},
  }

\ifCLASSINFOpdf
\else
\fi
\begin{document}
\title{Blind Identification of SFBC-OFDM Signals Based on the Central Limit Theorem}
\author{  Mingjun Gao, \IEEEmembership{Student Member,~IEEE,}
        Yongzhao Li, \IEEEmembership{Senior Member,~IEEE,}\\
        Octavia A. Dobre, \IEEEmembership{Senior Member,~IEEE,} and
        Naofal Al-Dhahir, \IEEEmembership{Fellow,~IEEE} % <-this % stops a space
%\thanks{}% <-this % stops a space
%\thanks{This work was supported in part by the National Natural Science Foundation of China (61771365), and the Natural Science Foundation of Shaanxi Province (2017JZ022), the 111 Project (B08038), the National Key Research and Development Program of China (2016YF-B1200202) and the Fundamental Research Funds for the Central Universities. The work of M. Gao was supported in part by the scholarship from China Scholarship Council. The work of O. A. Dobre was supported in part by the Natural Sciences and Engineering Research Council  (NSERC) of Canada through its Discovery program. The work of N. Al-Dhahir was made possible by NPRP grant number 8-627-2-260 from the Qatar National Research Fund (a member of Qatar Foundation).}
%\thanks{M. Gao and Y. Li are with the State Key Laboratory of Integrated Services Networks, Xidian University, Xi'an 710071, China. Corresponding author: Y. Li (e-mail: yzhli@xidian.edu.cn)}% <-this % stops a space
%\thanks{O. A. Dobre is with the Department of Electrical and Computer Engineering, Memorial University, St. John's, NL A1B 3X5, Canada.}
%\thanks{N. Al-Dhahir is with the Department of Electrical and Computer Engineering , University of Texas at Dallas, Richardson, TX 75080.}
%\thanks{This paper has been presented in part in the 2017 IEEE GLOBECOM.}
}

\maketitle

\begin{abstract}
Previous approaches for blind identification of space-frequency block codes (SFBC) do not perform well for short observation periods due to their inefficient utilization of frequency-domain redundancy. This paper proposes a hypothesis test (HT)-based algorithm and a support vector machine (SVM)-based  algorithm for SFBC signals identification over frequency-selective fading channels to exploit two-dimensional space-frequency domain redundancy. Based on the central limit theorem, space-domain redundancy is used to construct the cross-correlation function of the estimator and frequency-domain redundancy is incorporated in the construction of the statistics. The difference between the two proposed algorithms is that the HT-based algorithm constructs a chi-square statistic and employs an HT to make the decision, while the SVM-based algorithm constructs a non-central chi-square statistic with unknown mean as a strongly-distinguishable statistical feature and uses SVM to make the decision. Both algorithms do not require knowledge of the channel coefficients, modulation type or noise power, and the SVM-based algorithm does not require timing synchronization. Simulation results verify the superior performance of the proposed algorithms for short observation periods with comparable computational complexity to conventional algorithms, as well as their acceptable identification performance in the presence of transmission impairments.
\end{abstract}


% \fi
\IEEEpeerreviewmaketitle

\section{Introduction}

\IEEEPARstart{B}{lind} identification of communication signals' parameters of a transmitter from received signals without reference signals plays a vital role in many military and civilian applications. In military communication systems, the identified parameters are extremely important to carry out electronic warfare operations including surveillance, information decoding, and jamming signal design. In addition, software-defined and cognitive radios which are adopted in civilian applications also employ blind identification to sense signals and automatically adjust the design parameters of the transmitter\cite{Survey_Signal_Identification}. Recently, blind identification of multiple-input multiple-output (MIMO) or MIMO-orthogonal frequency division multiplexing (OFDM) signals has received considerable interest including enumeration of the number of transmit antennas \cite{AIC_MDL,HOM_TD_Nt_est,WME,My_paper_TWC1} and identification of space-time/frequency block codes (STBC/SFBC)  \cite{Likelihood_Based,correlator_function,Fourth_order_TC, Second_Order_cyclic,K_S_test,Classify_STBC_Over_FS,STBC_cyclic_2015_ICC,Blind_MIMO_OFDM,Blind_MIMO_OFDM_SM_AL,Identification_SM_AL_OFDM_cyclic,blind_SFBC,My_paper_Globecom,My_paper_TVT,My_paper_TWC1}.

Previous works on the identification of STBCs/SFBCs include references \cite{Likelihood_Based,correlator_function,Fourth_order_TC, Second_Order_cyclic,K_S_test,Classify_STBC_Over_FS,STBC_cyclic_2015_ICC,My_paper_TWC1} for single-carrier systems and references \cite{Blind_MIMO_OFDM,Blind_MIMO_OFDM_SM_AL,Identification_SM_AL_OFDM_cyclic,blind_SFBC,My_paper_Globecom,My_paper_TVT,My_paper_TWC1} for OFDM systems. Regarding the identification of STBCs for single-carrier systems, the reported algorithms can be divided into two types: likelihood-based \cite{Likelihood_Based} and feature-based \cite{correlator_function,Fourth_order_TC, Second_Order_cyclic,K_S_test,Classify_STBC_Over_FS,STBC_cyclic_2015_ICC,My_paper_TWC1} algorithms. The former uses the likelihood functions of the received signals to classify STBCs with different code rates. Reference \cite{My_paper_TWC1} quantifies the space-time/frequency redundancies as  features and employs an artificial neural network to distinguish between the features to jointly identify the number of transmit antennas and STBCs for both single-carrier and OFDM systems. The other feature-based methods detect the presence of the space-time redundancy at some specific time-lag locations by examining signal statistics or cyclic statistics. Most of these algorithms can not identify STBC/SFBC-OFDM signals since they do not work in the frequency-selective fading environment.
As for STBC-OFDM systems, such as WiFi \cite{STBC_on_80211}, references \cite{Blind_MIMO_OFDM,Blind_MIMO_OFDM_SM_AL,Identification_SM_AL_OFDM_cyclic} utilize the time-domain cross-correlation between adjacent OFDM symbols, i.e., the space-time redundancy, as a discriminating feature. Specifically, references \cite{Blind_MIMO_OFDM,Blind_MIMO_OFDM_SM_AL} use different cross-correlation functions, while reference \cite{Identification_SM_AL_OFDM_cyclic} employs a cyclic cross-correlation function with a specific time-lag over adjacent OFDM symbols. However, SFBC-OFDM, where the SFBC is employed over consecutive sub-carriers of an OFDM symbol, is preferable over STBC-OFDM for higher mobility applications, such as LTE \cite{sesia2009lte} and WiMAX \cite{IEEE802_16,STBC_4_ant}, since implementing the STBC over consecutive OFDM symbols is not effective due to the time-varying channels \cite{ICI_affects_OSTBC}. Hence, the time-domain cross-correlation between consecutive OFDM symbols does not exist any longer for SFBC-OFDM signals and the peaks of the cross-correlation function proposed in \cite{Blind_MIMO_OFDM,Blind_MIMO_OFDM_SM_AL,Identification_SM_AL_OFDM_cyclic} are difficult to detect. Therefore, blind identification algorithms of STBC-OFDM signals cannot be directly applied to SFBC-OFDM signals.

References \cite{blind_SFBC,My_paper_Globecom,My_paper_TVT,My_paper_TWC1} are the previous relevant works on the identification of SFBC-OFDM signals. Reference \cite{blind_SFBC} extends the idea of detecting the peak of the cross-correlation function with specific time lags between two receive antennas to the identification of SFBC-OFDM signals which only takes advantage of the space-domain redundancy. However, the frequency-domain redundancy is not utilized effectively which results in a negligible improvement of the performance when increasing the number of OFDM sub-carriers. Additionally, $N$ cross-correlation values are still calculated to determine the location of the peak ($N$ is the number of OFDM sub-carriers). To make use of the frequency-domain redundancy, we proposed to identify SFBC-OFDM signals by quantifying and distinguishing the frequency-domain redundancy of adjacent OFDM sub-carriers in \cite{My_paper_TVT,My_paper_TWC1}. However, the performance improvement is small since the probability of correctly identifying the SFBC signals converges rapidly with increasing $N$. Our prior work in \cite{My_paper_Globecom} does not consider multiple receive antenna pairs to improve the performance, and lacks the theoretical performance analysis of identifying SFBC signals.

In this paper, by exploiting the two-dimensional space-frequency domain redundancy, a hypothesis test (HT)-based blind identification algorithm and a support vector machine (SVM)-based blind identification algorithm for SFBC signals are proposed to improve the performance when increasing $N$ or for a small observation period over frequency-selective fading channels. Specifically, the space-domain redundancy is used for designing an estimator which is a cross-correlation function between antenna pairs. Furthermore, based on the central limit theorem (CLT), the frequency-domain redundancy is utilized by constructing the statistical features from the received signals on multiple OFDM sub-carriers. Regarding the utilization of the frequency-domain redundancy, {\bf{1)}} the first algorithm constructs a test statistic from multiple OFDM sub-carriers which follows a chi-square distribution for spatial multiplexing (SM) signals but not for SFBC signals. Then, an HT is proposed to make the decision; {\bf{2)}} the second algorithm is based on a strongly-distinguishable statistic which follows a non-central chi-square distribution with unknown mean for SM signals. Then, a trained SVM is used to identify SFBC signals. Both proposed algorithms can improve the identification performance as the number of OFDM sub-carriers increases, as well as provide satisfactory identification performance under frequency-selective fading with a shortened observation period, due to efficient utilization of the frequency-domain redundancy. In addition, both algorithms do not require \textit{a priori} knowledge of the signal parameters, such as channel coefficients, modulation type or noise power, and the SVM-based algorithm does not require timing synchronization. Furthermore, both algorithms have a satisfactory computational complexity and can be efficiently implemented with a parallel architecture.

The main contributions of this paper are the following:
\begin{itemize}
\item
The cross-correlation statistics between receive antenna pairs for SFBC signals are derived by utilizing the space-domain redundancy for signal type identification. Then, an HT-based identification algorithm of SFBC-OFDM signals is proposed to efficiently utilize the frequency-domain redundancy by constructing the test statistic from the received signals at consecutive OFDM sub-carriers.
\item
We derive analytical expressions for the probability of correctly identifying the SM and Alamouti (AL)-SFBC signals for the HT-based algorithm at any signal-to-noise ratio (SNR). 
\item
An SVM-based identification algorithm for SFBC-OFDM signals is proposed to improve the distinguishability of the discriminating feature between SM and SFBC signals and relax the requirement of \textit{a priori} knowledge of the timing synchronization by reconstructing the test statistic. Then, a trained SVM is used to make the decision. 
\item
The computational complexity is analyzed and shown to be satisfactory in comparison with the algorithms in \cite{blind_SFBC,My_paper_TVT}.
\item
Simulation results are presented to demonstrate the viability of the proposed algorithms with different design parameters and also in the presence of transmission impairments, including timing and frequency offsets, as well as Doppler effects.  
\end{itemize}

\begin{table*}
\centering
\caption{Notations.}
\label{notations}
\begin{tabular}{@{}cc|cc@{}}
\toprule
Notations                  & Descriptions              & Notations  &   Descriptions                                        \\ \midrule
${\left[  \cdot  \right]^ T }$    &  Transposition     & ${\left(  \cdot  \right)^ * }$ &    Complex conjugate                   \\
 $\left|  \cdot  \right|$            &  Absolute value of a number     &  ${\rm card}(\cdot)$ &   Cardinality for a set     \\           
$\lVert \cdot \rVert _F$           &  Frobenius norm        & $\ne$ &  Not equal sign                \\
$\Pr \left( B \right)$               &  Probability of the event $B$     &  ${\rm{E}}\left[  \cdot  \right]$ &  Statistical expectation     \\
\multirow{2}{*}{$\delta \left(  \cdot  \right)$}   &  Kronecker delta function where& \multirow{2}{*}{${\bf{I}}_m$} & \multirow{2}{*}{$m \times m$ identity matrix} \\[-1pt]
&$\delta \left(  0  \right)=1$ and is zero otherwise&&\\
$\rm tr \left( \cdot \right)$  &Trace of a matrix  & ${\rm diag}\left( \cdot \right)$ & Diagonal matrix\\
$e$ & Euler's constant&  $\rm{exp} \left( \cdot \right)$ & Exponential function\\
$\rm{log} \left( \cdot \right)$ & Logarithmic function & $\mathbb{N}$ &  Set of natural numbers\\ 
\multirow{2}{*}{${\cal N}\left( {\bf{0},\bf{I}} \right)$} &  \multirow{2}{*}{Standard normal distribution}& \multirow{2}{*}{$\chi _t^2$} & Central chi-squared distribution\\[-1pt]
&&&with $t$ degrees of freedom\\
\multirow{2}{*}{${d^{\left( i \right)}}$}& The symbol $d$ at the $i$-th trans- & \multirow{2}{*}{${d^{\left( i_1, i_2 \right)}}$} & The variable $d$ is dependent on the $i_1$-th\\[-1pt]
& mit or receive antenna&& and $i_2$-th transmit or receive antenna \\ \bottomrule
\end{tabular}
\end{table*}

This paper is organized as follows. In Section II, the signal model is introduced. Then, the HT-based algorithm and its theoretical performance analysis are presented in Section III. Next, the SVM-based algorithm is described in Section IV. The simulation results are presented in Section V. Finally, conclusions are drawn in Section VI. The summary of notations is presented in Table \ref{notations}.

%Standard notation is used throughout the paper. The superscripts ${\left[  \cdot  \right]^ T }$ and ${\left(  \cdot  \right)^ * }$ denote transposition and complex conjugate, respectively, $\left|  \cdot  \right|$ denotes the absolute value of a number. ${\rm card}(\cdot)$ denotes the cardinality for a set, $\lVert \cdot \rVert _F$ denotes the Frobenius norm, $\ne$ is the not equal sign. $\Pr \left( B \right)$ represents the probability of the event $B$, ${\rm{E}}\left[  \cdot  \right]$ indicates statistical expectation, and $\delta \left(  \cdot  \right)$ is the Kronecker delta function where $\delta \left(  0  \right)=1$ and is zero otherwise. $\bf{I}$ denotes the identity matrix, $\rm tr \left( \cdot \right)$ denotes the trace of a matrix, while ${\rm diag}\left( \cdot \right)$ denotes the diagonal matrix. In addition, $e$, $\rm{exp} \left( \cdot \right)$ and $\rm{log} \left( \cdot \right)$ denote Euler's constant, the exponential, and the logarithmic function, respectively. $\mathbb{N}$ denotes the set of natural numbers. ${\cal N}\left( {\bf{0},\bf{I}} \right)$ represents the standard normal distribution, $\chi _t^2$ denotes a central chi-squared distribution with $t$ degrees of freedom, ${d^{\left( i \right)}}$ is the symbol $d$ at the $i$-th transmit or receive antenna, and the notation ${d^{\left( i_1, i_2 \right)}}$ indicates that the variable $d$ is dependent on the $i_1$-th and $i_2$-th transmit or receive antenna.

\section{System Model}

We consider a MIMO-OFDM system with $N_t$ transmit antennas, $N_r$ ($N_r \ge 2$) receive antennas, $N$ sub-carriers and $\nu$ cyclic prefix samples. At the transmitter, the data symbols are drawn from an $M$-Phase Shift Keying (PSK) or  $M$-Quadrature Amplitude Modulation (QAM) signal constellation and parsed into data blocks, where each block ${{\bf{x}}_b} = {\left[ {{x_{b,0}}, \cdots ,{x_{b,{N_s} - 1}}} \right]^T}$ ($b \in {\mathbb{N}}$) consists of ${N_s}$ symbols. The SFBC encoder takes an $N_t \times L$ codeword matrix, denoted by ${\bf{C}}\left( {{{\bf{x}}_b}} \right)$, to span $L$ consecutive sub-carries in an OFDM symbol. In this paper, the codewords include SM, AL and two SFBCs with different code rates \cite{STBC_Tarokh} whose codeword matrices are given by
%\begin{gather}
% {{\bf{C}}^{{\rm{SM}}}} ( {{{\bf{x}}_b}}  ) = { [ {{x_{b,0}}, \cdots ,{x_{b,{N_t} - 1}}}  ]^T}  \label{eq1} \\
% {{\bf{C}}^{{\rm{AL}}}} ( {{{\bf{x}}_b}}  ) =  \left[ {\begin{array}{*{20}{c}}
%{{x_{b,0}}}&{ - x_{b,1}^ * }\\
%{{x_{b,1}}}&{x_{b,0}^ * }
%\end{array}}  \right]   . \label{eq2}
%\end{gather}
%\begin{gather}
% {{\bf{C}}^{{\rm{SM}}}} ( {{{\bf{x}}_b}}  ) = { [ {{x_{b,0}}, \cdots ,{x_{b,{N_t} - 1}}}  ]^T}  \label{eq1}\\
% {{\bf{C}}^{{\rm{AL}}}} ( {{{\bf{x}}_b}}  ) =  \left[ {\begin{array}{*{20}{c}}
%{{x_{b,0}}}&{ - x_{b,1}^ * } \label{eq2}\\
%{{x_{b,1}}}&{x_{b,0}^ * }
%\end{array}}  \right]  \\
%\mathbf{C}^{\text{SFBC}1}\left( \mathbf{x}_b \right) =\left[ \begin{matrix}
%	x_{b,0}&		-x_{b,1}&		-x_{b,2}&		-x_{b,3}&		x_{b,0}^{*}&		-x_{b,1}^{*}&		-x_{b,2}^{*}&		-x_{b,3}^{*}\\
%	x_{b,1}&		x_{b,0}&		x_{b,3}&		-x_{b,2}&		x_{b,1}^{*}&		x_{b,0}^{*}&		x_{b,3}^{*}&		-x_{b,2}^{*}\\
%	x_{b,2}&		-x_{b,3}&		x_{b,0}&		x_{b,1}&		x_{b,2}^{*}&		-x_{b,3}^{*}&		x_{b,0}^{*}&		x_{b,1}^{*}\\
%\end{matrix} \right]  \label{eq3}\\
%\mathbf{C}^{\text{SFBC}2}\left( \mathbf{x}_b \right) =\left[ \begin{matrix}
%	x_{b,0}&		-x_{b,1}^{*}&		\frac{x_{b,2}^{*}}{\sqrt{2}}&		\frac{x_{b,2}^{*}}{\sqrt{2}}\\
%	x_{b,1}&		x_{b,0}^{*}&		\frac{x_{b,2}^{*}}{\sqrt{2}}&		-\frac{x_{b,2}^{*}}{\sqrt{2}}\\
%	\frac{x_{b,2}}{\sqrt{2}}&		\frac{x_{b,2}}{\sqrt{2}}&		\frac{-x_{b,0}-x_{b,0}^{*}+x_{b,1}-x_{b,1}^{*}}{2}&		\frac{x_{b,1}+x_{b,1}^{*}+x_{b,0}-x_{b,0}^{*}}{2} \\
%\end{matrix} \right]   \label{eq4}.
%\end{gather}
\begin{equation}
 {{\bf{C}}^{{\rm{SM}}}} ( {{{\bf{x}}_b}}  ) = { [ {{x_{b,0}}, \cdots ,{x_{b,{N_t} - 1}}}  ]^T}  \label{eq1}
 \end{equation}
 \begin{equation}
 {{\bf{C}}^{{\rm{AL}}}} ( {{{\bf{x}}_b}}  ) =  \left[ {\begin{array}{*{20}{c}}
{{x_{b,0}}}&{ - x_{b,1}^ * } \label{eq2}\\
{{x_{b,1}}}&{x_{b,0}^ * }
\end{array}}  \right]  
\end{equation}
\begin{equation}
{{\bf{C}}^{\rm{SFBC1}}}({{\bf{x}}_b}) = {\left[ {\begin{array}{*{20}{c}}
{{x_{b,0}}}&{{x_{b,1}}}&{{x_{b,2}}}\\
{ - {x_{b,1}}}&{{x_{b,0}}}&{ - {x_{b,3}}}\\
{ - {x_{b,2}}}&{{x_{b,3}}}&{{x_{b,0}}}\\
{ - {x_{b,3}}}&{ - {x_{b,2}}}&{{x_{b,1}}}\\
{x_{b,0}^*}&{x_{b,1}^*}&{x_{b,2}^*}\\
{ - x_{b,1}^*}&{x_{b,0}^*}&{ - x_{b,3}^*}\\
{ - x_{b,2}^*}&{x_{b,3}^*}&{x_{b,0}^*}\\
{ - x_{b,3}^*}&{ - x_{b,2}^*}&{x_{b,1}^*}
\end{array}} \right]^T}  \label{eq3}
\end{equation}
\begin{equation}
\mathbf{C}^{\rm{SFBC2}}\left( \mathbf{x}_b \right) =\left[ \begin{matrix}
	x_{b,0}&		x_{b,1}&		\frac{x_{b,2}}{\sqrt{2}}\\
	-x_{b,1}^{*}&		x_{b,0}^{*}&		\frac{x_{b,2}}{\sqrt{2}}\\
	\frac{x_{b,2}^{*}}{\sqrt{2}}&		\frac{x_{b,2}^{*}}{\sqrt{2}}&		\frac{-x_{b,0}-x_{b,0}^{*}+x_{b,1}-x_{b,1}^{*}}{2}\\
	\frac{x_{b,2}^{*}}{\sqrt{2}}&		-\frac{x_{b,2}^{*}}{\sqrt{2}}&		\frac{x_{b,1}+x_{b,1}^{*}+x_{b,0}-x_{b,0}^{*}}{2}\\
\end{matrix} \right]^T   \label{eq4}.
\end{equation}
The symbol in the $i$-th row of ${\bf{C}}\left( {{{\bf{x}}_b}} \right)$ is transmitted from the $i$-th antenna. The symbols are input to $N$ consecutive OFDM sub-carriers of one block. Thus, the OFDM block is represented as
\begin{equation}
{\bf{S}}\left( {{{\bf{x}}_b}, \cdots ,{{\bf{x}}_{b + \frac{N}{L} - 1}}} \right) = \left[ {{\bf{C}}\left( {{{\bf{x}}_b}} \right), \cdots ,{\bf{C}}\left( {{{\bf{x}}_{b + \frac{N}{L} - 1}}} \right)} \right].  \label{eq5}
\end{equation}
Then, an $N$-point inverse fast Fourier transform (IFFT) converts this block into a time-domain block, and the last $\nu $ samples are appended as a cyclic prefix (CP).

At the receiver side, to simplify the derivations, we assume a perfect synchronizer at the beginning; however, we will analyze the sensitivity to model mismatches in Section V.\footnote{Blind synchronization can be achieved by utilizing the cyclostationarity of the received OFDM symbols \cite{Cyclic_of_OFDM,OFDM_parameters}. In addition, the SVM-based algorithm relaxes this assumption.} Then, the received OFDM symbol is converted to the frequency-domain via an $N$-point FFT after removing the CP. We can construct an $N_t$-dimensional transmitted signal vector which consists of one column of ${\bf{S}}\left( {{{\bf{x}}_b}, \cdots, {{\bf{x}}_{b + {N/L - 1}}}} \right)$, denoted by ${{\bf{s}}_k}(n) = {[ {s_k^{\left( 1 \right)}(n),\cdots,s_k^{\left( N_t \right)}(n)} ]^T}$, and an $N_r$-dimensional received signal vector, denoted by ${{\bf{y}}_k}(n) = {[ {y_k^{\left( 1 \right)}(n),\cdots,y_k^{\left( N_r \right)}(n)} ]^T}$ at the $k$-th ($1 \le k \le N$) sub-carrier of the $n$-th ($n \in {\mathbb{N}}$) OFDM symbol. The channel is assumed to be frequency-selective fading and the $k$-th subchannel is characterized by an $N_r \times N_t$ full-column rank matrix of fading coefficients denoted by
\begin{equation}
{{\bf{H}}_k} =  \left[ {\begin{array}{*{20}{c}}
{H_k^{( {1,1} )}}& \cdots &{H_k^{( {{N_t},1} )}}\\
 \vdots & \ddots & \vdots \\
{H_k^{( {1,{N_r}} )}}& \cdots &{H_k^{( {{N_t},{N_r}} )}}
\end{array}}  \right] \label{eq6}
\end{equation}
where ${H_k^{\left( {{i_1},{i_2}} \right)}}$ represents the channel coefficient between the $i_1$-th transmit and the $i_2$-th receive antenna. Then, the $n$-th  received signal at the $k$-th OFDM sub-carrier is expressed as
\begin{equation}
{{\bf{y}}_k}\left( n \right) = {{\bf{H}}_k} {{\bf{s}}_k}\left( n \right) + {{\bf{w}}_k}\left( n \right) \label{eq7}
\end{equation}
where the $N_r$-dimensional vector ${{\bf{w}}_k} = {[ {w_k^{\left( 1 \right)}\left( n \right),\cdots,w_k^{\left( N_r \right)}\left( n \right)} ]^T}$ represents the additive white Gaussian noise (AWGN) with zero mean and covariance $\sigma _w^2{{\bf{I}}_{N_r}}$ at the $k$-th OFDM sub-carrier.

%\subsection{Assumptions}
%
%\begin{assumption}
%The channel ${{\bf{H}}_k}$ is assumed to be of full-column rank and the channel gains remain constant over the observation period. The channel response and tranmitted signals are assumed statistically independent.
%\end{assumption}
%\begin{assumption}
% The data symbols are uncorrelated with ${\rm{E}}\left[ {{x_{b,m}}  {x_{b',m'}}} \right] = 0$ and ${\rm{E}}\left[ {{x_{b,m}}  x_{b',m'}^*} \right] = \sigma _s^2  \delta \left( {b - b'} \right)\delta \left( {m - m'} \right)$, where $\sigma _s^2$ is the transmit signal variance.
%\end{assumption}
%\begin{assumption}
% The transmitted signals and noise are uncorrelated, i.e., ${\rm{E}}\left[ {s_k^{\left( i \right)}\left( n \right)  w_{k'}^{\left( {i'} \right)}\left( {n'} \right)} \right] = 0$.
%\end{assumption}
%\begin{assumption}
%The noise in each channel is uncorrelated with that of the other channels: ${\rm{E}}\left[ {w_k^{\left( i \right)}\left( n \right)  w_{k'}^{\left( {i'} \right)}\left( {n'} \right)} \right] = \sigma _w^2 \delta \left( {k - k'} \right)\delta \left( {i - i'} \right)\delta \left( {n - n'} \right)$.
%\end{assumption}
 
\section{Proposed HT-Based Blind Identification Algorithm}

The correlation function for the single-antenna system has been investigated in \cite{correlation1,correlation2}. In this section, we design a cross-correlation function for multiple receive antennas to exploit the space-domain redundancy and propose an HT-based algorithm to take advantage of the frequency-domain redundancy. The frequency-domain redundancy among multiple consecutive OFDM sub-carriers can be formulated as a chi-square statistic for SM signals using the cross-correlation and the CLT. In addition, a threshold is employed to check the test statistic and make the decision. Moreover, the theoretical expressions of the probability of correctly identifying the SM and AL-SFBC signals are derived and analyzed. Furthermore, a decision tree is proposed to identify other SFBC signals.

\subsection{Cross-Correlation Function at the Receiver}

First, we define the cross-correlation function ${{R}^{\left( i_1, i_2 \right)}}\left( {k_1,k_2} \right)$ between the $k_1$-th OFDM sub-carrier at the $i_1$-th receive antenna and $k_2$-th OFDM sub-carrier at the $i_2$-th receive antenna as 
\begin{equation}
{{R}_C^{\left( i_1, i_2 \right) }}\left( {k_1,k_2} \right) = {\rm{E}}\left[ {y_{k_1}^{\left( i_1 \right)}(n) y_{k_2}^{\left( i_2 \right)}(n)} \right] \label{eq8}
\end{equation}
where $i_1 \ne i_2$ and $C $ denotes the SFBC, i.e., $C \in \{ \rm{SM}, \rm{AL}, \rm{SFBC1}, \rm{SFBC2} \}$. We can write the following expressions for the SFBC signals.

\subsubsection{SM-SFBC}

Assume that the data and noise are uncorrelated with ${\rm{E}}[ {s_k^{\left( i \right)}\left( n \right)  w_{k'}^{( {i'} )}\left( {n'} \right)} ] = 0$, the noises are independent with ${\rm{E}}[ {w_k^{\left( i \right)}\left( n \right)  w_{k'}^{( {i'} )}\left( {n'} \right)} ] = \sigma _w^2 \delta \left( {k - k'} \right)\delta \left( {i - i'} \right)\delta \left( {n - n'} \right)$, and the data symbols are uncorrelated with ${\rm{E}}[ {{x_{b,m}}  {x_{b',m'}}} ] = 0$ and ${\rm{E}}[ {{x_{b,m}}  x_{b',m'}^*} ] = \sigma _s^2  \delta \left( {b - b'} \right)\delta \left( {m - m'} \right)$, where $\sigma _s^2$ is the transmit signal variance. Without loss of generality, the index $n$ is omitted. Assume that the samples at the $k_1$-th and $k_2$-th ($k_1 \ne k_2$) OFDM sub-carriers over one transmission are ${{x_{b_1,0}}}$, ${{x_{b_1,1}}}$ and ${{x_{b_2,0}}}$, ${{x_{b_2,1}}}$, respectively. Based on \eqref{eq7} and \eqref{eq8}, we have
\begin{align}
{{R}_{\rm SM}^{\left( i_1, i_2 \right) }}\left( {k_1,k_2} \right) =& \   {\rm{E}}\left[ {H_{k_1}^{\left( {1,i_1} \right)}H_{k_2}^{\left( {1,i_2} \right)}s_{k_1}^{\left( 1 \right)}s_{k_2}^{\left( 1 \right)}} \right] + \notag \\
& \  {\rm{E}}\left[ {H_{k_1}^{\left( {1,i_1} \right)}H_{k_2}^{\left( {2,i_2} \right)}s_{k_1}^{\left( 1 \right)}s_{k_2}^{\left( 2 \right)}} \right] + \notag \\
& \  {\rm{E}}\left[ {H_{k_1}^{\left( {2,i_1} \right)}H_{k_2}^{\left( {1,i_2} \right)}s_{k_1}^{\left( 2 \right)}s_{k_2}^{\left( 1 \right)}} \right] + \notag \\
& \  {\rm{E}}\left[ {H_{k_1}^{\left( {2,i_1} \right)}H_{k_2}^{\left( {2,i_2} \right)}s_{k_1}^{\left( 2 \right)}s_{k_2}^{\left( 2 \right)}} \right] \notag \\
 =& \ {\rm{E}}\left[ {H_{k_1}^{\left( {1,i_1} \right)}H_{k_2}^{\left( {1,i_2} \right)}{x_{b_1,0}}{x_{b_2,0}}} \right] + \notag \\
 & \  {\rm{E}}\left[ {H_{k_1}^{\left( {1,i_1} \right)}H_{k_2}^{\left( {2,i_2} \right)}{x_{b_1,0}}{x_{b_2,1}}} \right] + \notag \\
& \  {\rm{E}}\left[ {H_{k_1}^{\left( {2,i_1} \right)}H_{k_2}^{\left( {1,i_2} \right)}{x_{b_1,1}}{x_{b_2,0}}} \right] + \notag \\
& \  {\rm{E}}\left[ {H_{k_1}^{\left( {2,i_1} \right)}H_{k_2}^{\left( {2,i_2} \right)}{x_{b_1,1}}{x_{b_2,1}}} \right]  = 0. \label{eq9}
\end{align}

\subsubsection{AL-SFBC}

 The samples at the $k$-th and $(k+1)$-th OFDM sub-carriers are denoted by ${x_{b,0}}$, $ - x_{b,1}^*$ and ${x_{b,1}}$, $x_{b,0}^*$, respectively. From \eqref{eq2}, \eqref{eq7} and \eqref{eq8}, the cross-correlation function of the received signals at two consecutive OFDM sub-carriers is
\begin{align}
{{R}_{\rm AL}^{\left( i_1, i_2 \right) }}\left( {k,k + 1} \right) = & \   {\rm{E}}\left[ {H_k^{\left( {1,i_1} \right)}H_{k + 1}^{\left( {1,i_2} \right)}{x_{b,0}}{x_{b,1}}} \right] + \notag \\
& \   {\rm{E}}\left[ {H_k^{\left( {1,i_1} \right)}H_{k + 1}^{\left( {2,i_2} \right)}{x_{b,0}}x_{b,0}^*} \right] - \notag \\
& \  {\rm{E}}\left[ {H_k^{\left( {2,i_1} \right)}H_{k + 1}^{\left( {1,i_2} \right)}{x_{b,1}}x_{b,1}^*} \right] - \notag \\ 
& \  {\rm{E}}\left[ {H_k^{\left( {2,i_1} \right)}H_{k + 1}^{\left( {2,i_2} \right)}x_{b,0}^*x_{b,1}^*} \right] \notag \\
 =  & \  \left( {H_k^{\left( {1,i_1} \right)}H_{k + 1}^{\left( {2,i_2} \right)} - H_k^{\left( {2,i_1} \right)}H_{k + 1}^{\left( {1,i_2} \right)}} \right)\sigma _s^2. \label{eq10}
\end{align}
Equation \eqref{eq10} shows that the cross-correlation is nonzero because each channel is statistically independent of the other channels.

\subsubsection{SFBC1}

From the codeword matrix of SFBC1, we have
\begin{align}
{R}_{\rm SFBC1}^{\left( i_1, i_2 \right) }\left( k,k+4 \right) = & \ ( H_{k}^{\left( 1,i_1 \right)}H_{k+4}^{\left( 2,i_2 \right)}+H_{k}^{\left( 2,i_1 \right)}H_{k+4}^{\left( 1,i_2 \right)} + \notag \\
& \  H_{k}^{\left( 3,i_1 \right)}H_{k+4}^{\left( 3,i_2 \right)} ) \sigma _{s}^{2} . \label{eq11}
\end{align}

\subsubsection{SFBC2}

Analogously, we have
\begin{align}
{R}_{\rm SFBC2}^{\left( i_1, i_2 \right) }\left( k,k+2 \right)= & \  ( H_{k}^{\left( 3,i_1 \right)}H_{k+2}^{\left( 1,i_2 \right)}+H_{k}^{\left( 3,i_1 \right)}H_{k+2}^{\left( 2,i_2 \right)} - \notag \\
& \  H_{k}^{\left( 1,i_1 \right)}H_{k+2}^{\left( 3,i_2 \right)}-H_{k}^{\left( 2,i_1 \right)}H_{k+2}^{\left( 3,i_2 \right)} ) \frac{\sigma _{s}^{2}}{2}. \label{eq12}
\end{align}

%{\color{blue}{In addition, we assume that the first received signal ${{y}_1}$ corresponds to the beginning of an SFBC block for simplifying the derivations, but relax this assumption in Section V.G.}}

\subsection{HT-Based Identification Algorithm of SM and AL-SFBC Signals}

Without loss of generality, we analyze the identification of AL versus SM signals in this section and the analysis of the other SFBCs is presented later, in Section III.D.  Define a set of receive antenna pairs with the cardinality $D = N_r(N_r-1)$ as
\begin{equation}
\Omega =\left\{ \left( i_1,i_2 \right) :i_1\ne i_2, 1 \le i_1 \le N_r, 1 \le i_2 \le N_r \right\}. \label{eq13}
\end{equation} 
For convenience, we simplify the form ${{ X}^{(i_1,i_2)} \left( {k_1,k_2} \right)}$ as ${ X\left( {k_1,k_2} \right)}$ unless otherwise stated. Then, the cross-correlation function estimator of the $i_1$-th and $i_2$-th receive antennas is given by
%\begin{equation}
\begin{align}
\hat{R}^{\left( i_1,i_2 \right)}\left( k_1,k_2 \right) & =\frac{1}{N_b}\sum_{n=1}^{N_b}{y_{k_1}^{\left( i_1 \right)}\left( n \right)  y_{k_2}^{\left( i_2 \right)}\left( n \right)}  \notag \\
& =R_{ C }\left( k_1,k_2 \right) +\epsilon \left( k_1,k_2 \right)  \label{eq14}
\end{align}
%\end{equation}
where $N_b$ is the number of received OFDM symbols, $\epsilon$ represents the estimation error which vanishes asymptotically as ${N_b} \to \infty $. Due to the error $\epsilon \left( {k_1,k_2} \right)$, the estimators ${{\hat {R}}}\left( {k_1,k_2} \right)$ are seldom exactly zero in practice for SM. To identify whether the received signals are AL or SM, we formulate the following HT problem
\begin{equation}
\begin{array}{l}
{{\bf{{\cal H}}}_0}:{{\hat {R}}}\left( {k,k + 1} \right) = \epsilon \left( {k,k + 1} \right) \\
{{\bf{{\cal H}}}_1}:{{\hat {R}}}\left( {k,k + 1} \right) = {{R}_{ { \rm{AL} } }}\left( {k,k + 1} \right) + \epsilon \left( {k,k + 1} \right)
\end{array}. \label{eq15}
\end{equation}
The estimator makes the decision that the signal type is SM under ${{\bf{{\cal H}}}_0}$ and AL under ${{\bf{{\cal H}}}_1}$. In this test, the distributions of $\epsilon \left( {k,k+1} \right)$ and ${{R}_{ { \rm{AL} } }}\left( {k,k + 1} \right)$ are required for decision. However, the statistical distributions are unknown at the receiver. Therefore, analyzing these distributions is the key to solve the problem, which we discuss next.

First, we obtain the $2D \times 1$ vectors $\mathbf{{r}}\left( k_1,k_2 \right)$ and ${\boldsymbol{\epsilon }}\left( k_1,k_2 \right)$ by stacking all the real and imaginary parts of the estimators and errors between the receive antenna pairs in $\Omega$ as follows
\begin{subequations} \label{eq16}
\begin{align}
\mathbf{{r}}\left( k_1,k_2 \right) &=\left[ \begin{array}{c}
	\Re \left\{ \hat{R}^{\left( 1,2 \right)}\left( k_1,k_2 \right) \right\}\\
	\vdots\\
	\Re \left\{ \hat{R}^{\left( i_1,i_2 \right)}\left( k_1,k_2 \right) \right\}\\
	\vdots\\
	\Im \left\{ \hat{R}^{\left( i_1,i_2 \right)}\left( k_1,k_2 \right) \right\}\\
	\vdots\\
\end{array} \right]  \\ 
{\boldsymbol{\epsilon }}\left( k_1,k_2 \right) &=\left[ \begin{array}{c}
	\Re \left\{ \epsilon ^{\left( 1,2 \right)}\left( k_1,k_2 \right) \right\}\\
	\vdots\\
	\Re \left\{ \epsilon ^{\left( i_1,i_2 \right)}\left( k_1,k_2 \right) \right\}\\
	\vdots\\
	\Im \left\{ \epsilon ^{\left( i_1,i_2 \right)}\left( k_1,k_2 \right) \right\}\\
	\vdots\\
\end{array} \right]. 
\end{align}
\end{subequations}
It is unnecessary to know the distribution of ${\boldsymbol{\epsilon }}\left( k_1,k_2 \right)$. Here, we assume that the errors of different estimators, i.e., ${\boldsymbol{\epsilon }}\left( k_1,k_2 \right)$, are independent and identically distributed random variables for different sub-carrier pairs. This is a reasonable assumption since the inputs of estimators are independent and the same type of signals. Therefore, ${\boldsymbol{\epsilon }}\left( k_1,k_2 \right)$ can be modeled as an independent zero-mean random vector with covariance matrix $\mathbf{\Psi }$. 

For {\bf{SM}} (under hypothesis ${{\bf{{\cal H}}}_0}$), according to the CLT, a group of vectors denoted by
\begin{equation}
\mathbf{u}_i=\mathbf{\Psi }^{-\frac{1}{2}}\mathbf{v}_i, \quad {i = 0, \cdots ,G-1 }  \label{eq17}
\end{equation}
follows an asymptotically standard normal distribution, i.e., $\mathbf{u }_i\rightarrow \mathcal{N}\left( {\bf{0}},{\mathbf{I}}_{2D} \right) $, for SM signals if $N' = N/G$ is a large number, where $G$ is the number of the vectors in the group and the vector $\mathbf{v}_i$ is given by
\begin{equation}
\mathbf{v }_i=\frac{1}{\sqrt{N'/2}}\sum_{j=iN'/2+1}^{(i+1)N'/2}{\mathbf{{r}}\left( 2j-1,2j \right)} . \label{eq18}
\end{equation} 
Moreover, the covariance matrix of the error vector $\boldsymbol{\epsilon }$ can be estimated as follows 
\begin{equation}
\mathbf{\hat{\Psi}}=\frac{1}{N-3}\sum_{k=1}^{N-2}{{\mathbf{I}}_{2D} \cdot \left[ {{\bf{ r}}\left( {k,k+2} \right) \circ {\bf{ r}}\left( {k,k+2} \right)} \right]}  \label{eq19}
\end{equation}
where $\cdot$ denotes the matrix multiplication and $\circ$ denotes the Hadamard product operation \cite{Had_Prod}. The Hadamard product guarantees a positive-definite $\mathbf{\hat{\Psi}}$. 
It is worth noting that we do not know which type of signal is received, and thus, use $\mathbf{{r}}\left( k,k+2 \right)$ to estimate $\mathbf{\hat{\Psi}}$. This is because $\mathbf{{r}}\left( k,k+2 \right)$ has the same distribution as $\boldsymbol{{\epsilon}}\left( k ,k+1 \right)$ regardless of the hypothesis. From \eqref{eq8}, we can easily show that ${{R}_{\rm AL}}\left( {k,k+2} \right) = 0$. Hence, each element of $\mathbf{{r}}\left( k,k+2 \right)$ can be expressed as follows
\begin{subequations}  \label{eq_ad1}
\begin{align}
{\rm Under}\  {{\bf{{\cal H}}}_0}:{{\hat {R}}}\left( {k,k + 2} \right) &= \epsilon \left( {k,k + 2} \right)  \\
{\rm Under}\  {{\bf{{\cal H}}}_1}:{{\hat {R}}}\left( {k,k + 2} \right) &= {{R}_{ { \rm{AL} } }}\left( {k,k + 2} \right) + \epsilon \left( {k,k + 2} \right)  \notag \\
                                                                                                    &= \epsilon \left( {k,k + 2} \right). 
\end{align}
\end{subequations} 
The error $\boldsymbol{{\epsilon}}\left( k ,k+2 \right)$ has the same distribution as $\boldsymbol{{\epsilon}}\left( k ,k+1 \right)$ as we discussed previously.
Then, we construct the following test statistic
\begin{equation}
{\cal U} = \sum\limits_{i = 0}^{G  - 1} {{\bf{v}}_i^T {\bf{\hat{\Psi }}}\!^{ - 1} {\bf{v}}_i} . \label{eq20}
\end{equation} 
Hence, the test statistic ${\cal U}=  \sum\nolimits_{i = 0}^{G  - 1}{{\bf{u}}_i^T {\bf{u}}_i}$ asymptotically follows a chi-square distribution with $q=2DG$ degrees of freedom, i.e., ${\cal U} \to \chi _{q}^2$.

\begin{figure}
  \centering
  \includegraphics[width=0.5\textwidth]{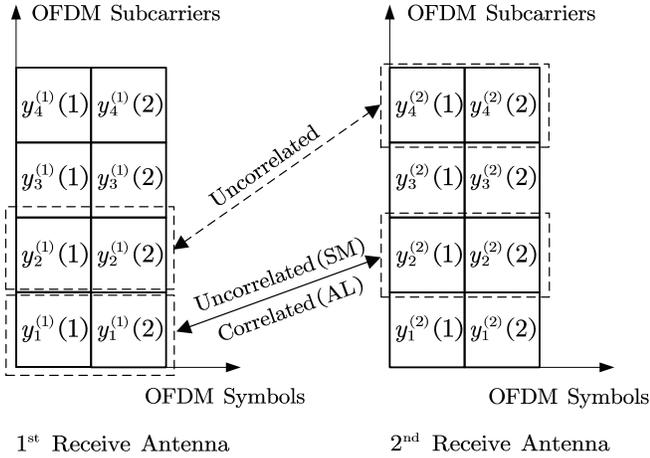}\\
  \caption{Cross-correlation between consecutive OFDM sub-carriers at the receiver. }\label{fig0}
\end{figure}

For {\bf AL-SFBC} (under hypothesis ${{\bf{{\cal H}}}_1}$), as shown in Fig. \ref{fig0}, since the signals at the $k$ and $(k+2)$ OFDM sub-carriers are uncorrelated and different from those at the $(2j-1)$ and $2j$ OFDM sub-carriers, $\hat {\bf{\Psi }}$ is not the covariance matrix of the vector $\mathbf{{r}}\left( 2j-1,2j \right)$ based on \eqref{eq15} and \eqref{eq19}. Therefore, ${\cal U}$ does not follow the standard chi-square distribution.

Accordingly, this observation allows us to design a detector threshold $\eta  $ which yields the desired probability of false alarm, ${{\rm Pr}_f}$, i.e., ${{\rm Pr}_f} = \Pr \left( {\left. {{{\bf{{\cal H}}}_1}} \right|{{\bf{{\cal H}}}_0}} \right) = \Pr \left( {{\cal U} \ge \eta  } \right)$.  Then, using  the cumulative distribution function (CDF) expression of the chi-square distribution, we find that
\begin{equation}
\Pr \left( {{\cal U} < \eta } \right) = \frac{\gamma \left( q/2,\eta /2 \right)}{\varGamma \left( q/2 \right)}  \label{eq21}
\end{equation}
where $\varGamma  \left( \cdot \right)$ is the Gamma function given by
\begin{equation}
\varGamma \left( m \right) =\left( m-1 \right) !  \label{eq22}
\end{equation}
and $\gamma \left( \cdot \right) $ is the lower incomplete Gamma function \cite{probability_handbook} given by
\begin{equation}
\gamma \left( \alpha,\beta \right) =\int_0^\beta{t^{\alpha-1}e^{-t}\text{d}t}.  \label{eq23}
\end{equation}
Since ${{\rm Pr}_f} = 1 - \Pr \left( {{\cal U} < \eta } \right)$, the threshold $\eta $ is calculated for a given ${{\rm Pr}_f}$ using the expression
\begin{equation}
\gamma \left( q/2,\eta /2 \right) =\left( q/2-1 \right) !\left( 1-{\rm Pr}_f \right)   . \label{eq24}
\end{equation} 
The threshold $\eta$ cannot be expressed in a closed-form since \eqref{eq24} is a nonlinear equation but can be numerically calculated by the bisection method \cite{numerical_analysis}. Then, if ${{\cal U} \ge \eta }$, the received signals are estimated as AL signals; otherwise, they are estimated as SM signals.

 For clarity, the main steps of the proposed algorithm are summarized as follows .
 \begin{algorithm}
 \caption{}
 \begin{algorithmic}[1]
 \renewcommand{\algorithmicrequire}{\textbf{Input: }}
 \renewcommand{\algorithmicensure}{\textbf{Output:}}
 \REQUIRE The observed synchronized sequence $\bf{y}$.
 \ENSURE  SFBC.
 \STATE{ Construct the stacked vectors $\mathbf{{r}}\left( 2j-1,2j \right)$, $j = 1, 2, \cdots , N/2$, and $\mathbf{{r}}\left( k,k+2 \right)$, $k = 1,2, \cdots ,N-1$, using \eqref{eq16}.}
 \STATE{ Compute the vectors ${\mathbf{v}}_i$ using \eqref{eq18}.} 
 \STATE{ Compute the covariance matrix $\mathbf{\hat{\Psi}'}$ using \eqref{eq19}. } 
 \STATE{ Construct the test statistic ${\cal U}$ using \eqref{eq20}.}
 \STATE{ Compute the threshold $\eta $ by calculating \eqref{eq24} via the bisection method.}
   \IF{${{\cal U} \ge \eta }$}
    \STATE {the AL-OFDM signal is declared present (${{\bf{{\cal H}}}_1}$ true).}
  \ELSE
    \STATE {the SM-OFDM signal is declared present (${{\bf{{\cal H}}}_0}$ true).}
  \ENDIF
\RETURN {SFBC.}
 \end{algorithmic}
 \end{algorithm}
 
 \subsection{Theoretical Performance Analysis for Identification of SM and AL-SFBC Signals}

{\bf Under hypothesis ${{\bf{{\cal H}}}_0}$}, as described previously, if $\cal U < \eta$, the SM signals are declared present. For a certain threshold $\eta$, the probability of correctly identifying the SM signals is determined as \cite{probability_handbook}
\begin{align}
{\rm Pr}({\rm SM|SM}) &= 1- {\rm Pr}_f   \notag \\
&= 1 - \exp \left( -\frac{\eta}{2} \right) \sum_{m=1}^{q/2}{\frac{1}{\left( m-1 \right) !}\left( \frac{\eta}{2} \right) ^{m-1}}. \label{eq25}
\end{align}

{\bf Under hypothesis ${{\bf{{\cal H}}}_1}$}, the probability of correctly identifying the AL signals is ${\rm Pr}({\rm AL|AL}) = {\rm Pr}({{\cal U \ge \eta} |{\cal H}_1})$. Without loss of generality, we analyze the simplest case here, namely, $\Omega = \{ (1,2)\}$ and $G=1$.  From \eqref{eq16}-\eqref{eq18}, the vector $\bf u$ is given by
\begin{equation}
{\bf{u}} = {\bf{\Psi }}^{ - 1/2}   \frac{1}{{\sqrt {N/2} }}\sum\limits_{j = 1}^{N/2} {\left[ {\begin{array}{*{20}c}
   {\Re \left\{ {\hat R\left( {2j - 1,2j} \right)} \right\}}  \\
   {\Im \left\{ {\hat R\left( {2j - 1,2j} \right)} \right\}}  \\
\end{array}} \right]}.  \label{eq26}
\end{equation}

\emph{Proposition 1:} Given the channel coefficients and denoting the vector ${{\bf{H}}_{k }^{(i)} }$ as the $i$-th row of ${\bf H}_k$ at the $k$-th OFDM sub-carrier, the covariance matrix $\mathbf{{\Psi}} = \sigma _{\epsilon}^{2}{\mathbf{I}}_{2}$, where $\sigma _{\epsilon}^{2}$ is given by
\begin{align}
\sigma _\epsilon ^2  =& \  \frac{{\sigma _s^4 }}{{2N_b }}\left\| {{\bf{H}}_{k_1 }^{(1)} } \right\|_F^2 \left\| {{\bf{H}}_{k_2 }^{(2)} } \right\|_F^2 + \notag \\  
&\  \frac{{\sigma _s^2 \sigma _w^2 }}{{2N_b }}\left( {\left\| {{\bf{H}}_{k_1 }^{(1)} } \right\|_F^2  + \left\| {{\bf{H}}_{k_2 }^{(2)} } \right\|_F^2 } \right) + \frac{{\sigma _w^4 }}{{2N_b }}. \label{eq27}
\end{align}

\emph{Proof:} See Appendix A.

Then, using \eqref{eq14} and \eqref{eq27}, $\bf u$ can be decomposed as follows
\begin{align}
{\bf{u}} =& \  \frac{1}{{\sqrt {N/2}  \cdot \sigma _\epsilon  }}\sum\limits_{j = 1}^{N/2} {\left[ {\begin{array}{*{20}c}
   {\Re \left\{ {R_{{\rm{AL}}} \left( {2j - 1,2j} \right)} \right\}}  \\
   {\Im \left\{ {R_{{\rm{AL}}} \left( {2j - 1,2j} \right)} \right\}}  \\
\end{array}} \right]}  + \notag \\
& \  \frac{1}{{\sqrt {N/2}  \cdot \sigma _\epsilon  }}\sum\limits_{j = 1}^{N/2} {\left[ {\begin{array}{*{20}c}
   {\Re \left\{ {\epsilon \left( {2j - 1,2j} \right)} \right\}}  \\
   {\Im \left\{ {\epsilon \left( {2j - 1,2j} \right)} \right\}}  \\
\end{array}} \right]} .  \label{eq28}
\end{align}
$\cal U$ is given by
\begin{equation}
{\cal U} = {\bf u}^T{\bf u}= a_1^2  + a_2^2  + a_1X_1  + a_2X_2  + X_1^2  + X_2^2  \label{eq29}
\end{equation}
where the two independent random variables $X_1$ and $X_2$ are, respectively, given by
\begin{subequations}  \label{eq30}
\begin{align}
&X_1 = \frac{1}{{\sqrt {N/2}  \cdot \sigma _\epsilon  }}\sum\limits_{j = 1}^{N/2} {\Re \left\{ {\epsilon \left( {2j - 1,2j} \right)} \right\}}  \\
&X_2 = \frac{1}{{\sqrt {N/2}  \cdot \sigma _\epsilon  }}\sum\limits_{j = 1}^{N/2} {\Im \left\{ {\epsilon \left( {2j - 1,2j} \right)} \right\}}
\end{align}
\end{subequations}
and they both asymptotically follow a standard normal distribution according to \eqref{eq17}, i.e., $X_1 \rightarrow \mathcal{N}\left( 0,1 \right)$ and $X_2 \rightarrow \mathcal{N}\left( 0,1 \right)$. Furthermore, the coefficients $a_1$ and $a_2$ are, respectively, given by
\begin{subequations}  \label{eq31}
\begin{align}
&a_1 = \frac{1}{{\sqrt {N/2}  \cdot \sigma _\epsilon  }}\sum\limits_{j = 1}^{N/2} {\Re \left\{ {R_{{\rm{AL}}} \left( {2j - 1,2j} \right)} \right\}}  \\
&a_2 = \frac{1}{{\sqrt {N/2}  \cdot \sigma _\epsilon  }}\sum\limits_{j = 1}^{N/2} {\Im \left\{ {R_{{\rm{AL}}} \left( {2j - 1,2j} \right)} \right\}}. 
\end{align}
\end{subequations}

\emph{Proposition 2:}  Given a real constant $\beta $ and a normally distributed random variable $X$ with CDF \cite{probability_handbook}
\begin{equation}
F_X\left( x \right) =1- \frac{1}{2} \text{erfc} \left( \frac{x}{\sqrt{2}} \right)  \label{eq32}
\end{equation}
where ${\rm erfc}(\cdot)$ is the complementary error function defined as
\begin{equation}
{\rm{erfc}}\left( \alpha  \right) = \frac{2}{{\sqrt {\pi}  }}\int_x^\infty  {e^{ - t^2 } {\rm{d}}t}   \label{eq33}
\end{equation}
the CDF of the random variable $Y=\beta X+X^2$ is as provided in \eqref{eq34}

\newcounter{TempEqCnt}
\setcounter{TempEqCnt}{\value{equation}}
\setcounter{equation}{34}
\begin{figure*}[ht]
  \normalsize

%  \vspace*{4pt}
\begin{equation}
F_Y \left( y \right) = \left\{ {\begin{array}{*{20}c}
   {\frac{1}{2}\left[ {\rm{erfc}}\left( {\frac{{ - \sqrt {y + \left( {\beta /2} \right)^2 }  - \beta /2}}{{\sqrt 2 }}} \right) - {{\rm{erfc}}\left( {\frac{{\sqrt {y + \left( {\beta /2} \right)^2 }  - \beta /2}}{{\sqrt 2 }}} \right)  } \right],} & {y \ge  - \frac{{\beta ^2 }}{4}}  \\
   {0,} & {y <  - \frac{{\beta ^2 }}{4}}  \\
\end{array}} \right. .   \label{eq34}
\end{equation}
\begin{subequations} \label{eq35}
\begin{align}
&F_{Y_1 } \left( {y_1 } \right) = \left\{ {\begin{array}{*{20}c}
   {\frac{1}{2}\left[ {\rm{erfc}}\left( {\frac{{ - \sqrt {y_1  + \left( {a_1/2} \right)^2 }  - a_1/2}}{{\sqrt 2 }}} \right) - {{\rm{erfc}}\left( {\frac{{\sqrt {y_1  + \left( {a_1/2} \right)^2 }  - a_1/2}}{{\sqrt 2 }}} \right)  } \right],} & {y_1  \ge  - \frac{{a_1^2 }}{4}}  \\
   {0,} & {y_1  <  - \frac{{a_1^2 }}{4}}  \\
\end{array}} \right.  \\
&F_{Y_2 } \left( {y_2 } \right) = \left\{ {\begin{array}{*{20}c}
   {\frac{1}{2}\left[ {\rm{erfc}}\left( {\frac{{ - \sqrt {y_2  + \left( {a_2/2} \right)^2 }  - a_2/2}}{{\sqrt 2 }}} \right) - {{\rm{erfc}}\left( {\frac{{\sqrt {y_2  + \left( {a_2/2} \right)^2 }  - a_2/2}}{{\sqrt 2 }}} \right) } \right],} & {y_2  \ge  - \frac{{a_2^2 }}{4}}  \\
   {0,} & {y_2  <  - \frac{{a_2^2 }}{4}}  \\
\end{array}} \right. . 
\end{align}
\end{subequations}
    \hrulefill
\end{figure*}

\emph{Proof:} See Appendix B.

Subsequently, two random variables $Y_1=a_1X_1+X_1^2$ and $Y_2=a_2X_2+X_2^2$ have the CDFs given as in \eqref{eq35}, respectively. 

Denote $Z = Y_1 + Y_2$, the CDF of $Z$ is
\begin{equation}
F_Z \left( z \right) = \int_{  -a_2^2/4 }^\infty  {F_{Y_1 } \left( {z - y_2 } \right)\,} {\rm{d}}F_{Y_2 } \left( {y_2 } \right) . \label{eq36}
\end{equation}
%\emph{Proposition 3:} With the CDFs of the random variables $Y_1$ and $Y_2$ defined in \eqref{eq35}, the CDF of the random variable $Z = Y_1 + Y_2$ is 
%\begin{equation}
%F_Z \left( z \right) = \int_{  -a_2^2/4 }^\infty  {F_{Y_1 } \left( {z - y_2 } \right)\,} {\rm{d}}F_{Y_2 } \left( {y_2 } \right) . \label{eq36}
%\end{equation}
%
%\emph{Proof:} See Appendix C.
Finally, the probability of correctly identifying the AL signals is 
\begin{equation}
{\rm Pr}({\rm AL|AL}) = 1- \int_{ -a_2^2/4 }^\infty  {F_{Y_1 } \left( {\eta - a_1^2 - a_2^2 - y_2 } \right)\,} {\rm{d}}F_{Y_2 } \left( {y_2 } \right) . \label{eq37}
\end{equation} 
Unfortunately, a closed-form expression for ${\rm Pr}({\rm AL|AL})$ does not exist. However, we compute ${\rm Pr}({\rm AL|AL})$ by using a numerical integration method such as the Riemann sum \cite{numerical_analysis}. Regarding the infinite upper limit of the integral in \eqref{eq37}, we can choose a big number as the upper limit since ${\rm{d}}F_{Y_2 } \left( {y_2 } \right) / {\rm{d}}y_2 $ quickly converges to zero when increasing $y_2$.  

For a general $\bf r$ having a large $\Omega $, $\cal U$ has the following more complicated expression
\begin{align}
{\cal U} = & \  a_1^2  + a_2^2 + \cdots   + a_{q}^2 + a_1X_1  + a_2X_2  + \cdots +  \notag \\
 & \ a_{q}X_{q} + X_1^2  + X_2^2 + \cdots +  X_{q}^2.  \label{eq38}
\end{align}
The probability of correctly identifying the AL signal can be expressed as a multiple integral which can be numerically evaluated using a numerical method as we previously described.

\subsection{Decision Tree for Identification of Three-Antenna SFBCs}

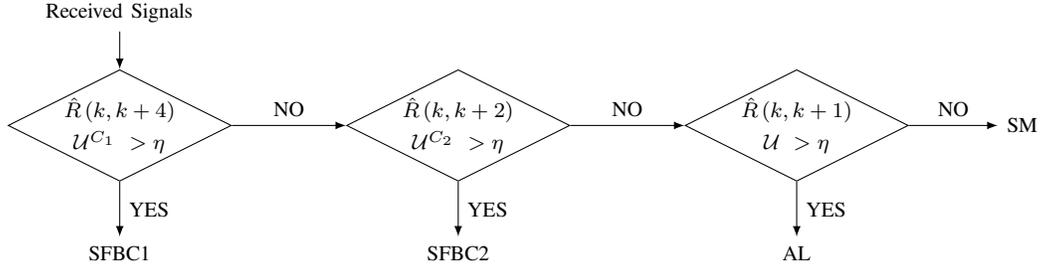
\begin{figure*}[ht]%* is used because without it the figure does not appear
\centering
\begin{tikzpicture}%[level 1/.style={sibling distance=4.5cm},level 2/.style={sibling distance=1.5cm},level 3/.style={sibling distance=2cm}]
\node (RS) at (0,0) {\footnotesize Received Signals};
\node [decision] (decide1)  at  (0,-1.5) {\footnotesize $\hat{R}\left( k,k+4 \right) $\\ \footnotesize ${\cal U}^{C_1}\  > \eta$};
\node  (SFBC1)   at  (0,-3.2)  {\footnotesize SFBC1};
\node [decision] (decide2)   at  (4.5,-1.5)  {\footnotesize $\hat{R}\left( k,k+2 \right) $\\ \footnotesize ${\cal U}^{C_2}\   > \eta$};
\node  (SFBC2)    at  (4.5,-3.2) {\footnotesize SFBC2};
\node [decision] (decide3)   at  (9,-1.5) {\footnotesize $\hat{R}\left( k,k+1 \right) $\\ \footnotesize ${\cal U}\ > \eta$};
\node  (AL)    at  (9,-3.2) {\footnotesize AL};
\node  (SM)   at  (12,-1.5)  {\footnotesize SM};

\draw [line] (RS) --  (decide1);
\draw [line] (decide1) -- node [right] {\footnotesize YES} (SFBC1);
\draw [line] (decide1) -- node [above] {\footnotesize NO} (decide2);
\draw [line] (decide2) -- node [right] {\footnotesize YES} (SFBC2);
\draw [line] (decide2) -- node [above] {\footnotesize NO} (decide3);
\draw [line] (decide3) -- node [right] {\footnotesize YES} (AL);
\draw [line] (decide3) -- node [above] {\footnotesize NO} (SM);	
%\path [line] (decide1) -| node [near start,above=-1.2mm] {\footnotesize YES} (SFBC2);
%\path [line] (decide1) -| node [near start,above=-1.2mm] {\footnotesize NO} (decide2);
%\path [line] (decide2) -| node [near start,above=-1.2mm] {\footnotesize YES} (SFBC1);
%\path [line] (decide2) -| node [near start,above=-1.2mm] {\footnotesize NO} (decide3);
%\path [line] (decide3) -| node [near start,above=-1.2mm] {\footnotesize YES} (AL);
%\path [line] (decide3) -| node [near start,above=-1.2mm] {\footnotesize NO} (SM);	
\end{tikzpicture}
\caption{Decision tree for the identification of SFBC signals.}
\label{Des_tree}

\end{figure*}

To identify the SFBC $C \in \{ \rm{SM}, \rm{AL}, \rm{SFBC1}, \rm{SFBC2} \}$, the previously described discriminating features are used with a decision tree classification algorithm, which is presented in Fig. \ref{Des_tree}. At the top-level node, the cross-correlation function estimator $\hat{R}\left( k,k+4 \right) $ is used to discriminate between $\rm SFBC1$ and the code $C_1 \in \{ \rm{SM}, \rm{AL}, \rm{SFBC2} \}$ based on the test statistic ${\cal U}^{C_1}$ and the threshold $\eta$ since the signals at the $k$ and $(k+4)$ sub-carriers are uncorrelated for $C_1$, i.e., ${R}_{C_1}\left( k,k+4 \right) = 0$. Similarly at the middle level node, $\hat{R}\left( k,k+2 \right) $ is used to discriminate between $\rm SFBC2$ and the code $C_2 \in \{ \rm{SM}, \rm{AL} \}$ based on the test statistic ${\cal U}^{C_2}$ and the same $\eta$. Here, the signals at the $k$ and $(k+2)$ sub-carriers are uncorrelated for $C_2$, i.e., ${R}_{C_2}\left( k,k+2 \right) = 0$. Finally, at the bottom level node, $\hat{R}\left( k,k+1 \right) $ is used as described previously. Here, the derivation of ${\cal U}^{C_1}$ and ${\cal U}^{C_2}$ are somehow tedious and are given in Appendix C. 

In particular, we can fix the probabilities of false alarm ${\rm Pr}_f$ for all the nodes of the decision tree in Fig. \ref{Des_tree}. Hence, the three nodes in the decision tree have identical $\eta$, which is calculated by solving \eqref{eq24}. This is because the test statistics ${{\cal U}^{{C_1}}}$, ${{\cal U}^{{C_2}}}$ and $\cal U$ follow the same distribution, namely, a chi-square distribution with $q$ degrees of freedom, for $C_1$, $C_2$ and SM, respectively. Different $\eta$ caused by different probabilities of false alarm indeed affect the performance. At the top-level node, a smaller $\eta$ leads to improved performance of identifying $\rm SFBC1$ but degrades the identification performance of $C_1$. A similar situation happens at the middle-level and bottom-level nodes.

\section{Proposed SVM-Based Blind Identification Algorithm}

Since the synchronization error in the time domain incurs a phase rotation in the frequency domain for OFDM signals \cite{MIMO-OFDM_with_MATLAB}, we propose an SVM-based algorithm to relax our assumption of perfect synchronization. After restructuring the statistic which follows a non-central chi-square distribution with an unknown mean for SM signals as a strongly-distinguishable statistical feature, a trained SVM is employed to classify different SFBC signals. Without loss of generality, we analyze the SM and AL signals in this section. The other SFBCs can be identified by using the same decision tree described in the previous section.

We construct new vectors ${\bf{ t}}\left( {k_1 ,k_2 } \right) $ and $\left|{\boldsymbol{\epsilon }}\left( k_1,k_2 \right) \right|$ by calculating the absolute value of each element of $\mathbf{{r}}\left( k_1,k_2 \right) $ and ${\boldsymbol{\epsilon }}\left( k_1,k_2 \right)$, respectively, as follows 
\begin{subequations}\label{eq39}
\begin{align}
{\bf{ t}}\left( {k_1 ,k_2 } \right) &= \left[ {\begin{array}{*{20}c}
    \vdots   \\
   {\left| {\Re \left\{ {\hat R^{\left( {i_1,i_2} \right)} \left( {k_1 ,k_2 } \right)} \right\}} \right|}  \\
    \vdots   \\
   {\left| {\Im \left\{ {\hat R^{\left( {i_1 ,i_2 } \right)} \left( {k_1 ,k_2 } \right)} \right\}} \right|}  \\
    \vdots   \\
\end{array}} \right],  \\ 
 \left|{\boldsymbol{\epsilon }}\left( k_1,k_2 \right) \right| &=\left[ \begin{array}{c}
       	\vdots\\
	{\left|\Re \left\{ \epsilon ^{\left( i_1,i_2 \right)}\left( k_1,k_2 \right) \right\} \right|}\\
	\vdots\\
	{\left|\Im \left\{ \epsilon ^{\left( i_1,i_2 \right)}\left( k_1,k_2 \right) \right\} \right|}\\
	\vdots \\
\end{array} \right] 
\end{align}
\end{subequations}
which are not affected by a phase rotation. Assume that ${\boldsymbol{\mu }}$ and $\mathbf{\Phi }$ are the mean vector and covariance matrix of the vector $\left|{\boldsymbol{\epsilon }}\left( k_1,k_2 \right) \right|$, respectively. 

For {\bf{SM}}, according to the CLT, a vector defined as
\begin{equation}
\mathbf{p}=\mathbf{\Phi }^{-\frac{1}{2}}\mathbf{q} \label{eq40}
\end{equation}
follows an asymptotically standard normal distribution, i.e., $\mathbf{p }\rightarrow \mathcal{N}\left( {\bf{0}},{\mathbf{I}}_{2D} \right) $, for SM signals, where the vector $\mathbf{q}$ is given by
\begin{equation}
\mathbf{q}=\frac{1}{\sqrt{N/2}}\sum_{j=1}^{N/2}{\left[ \mathbf{t}\left( 2j-1,2j \right) -{\boldsymbol{\mu }} \right]}. \label{eq41}
\end{equation}
Furthermore, the mean vector ${\boldsymbol{\mu }}$ and covariance matrix $\mathbf{\Phi }$ of $\left| \boldsymbol{\epsilon } \right|$ can be estimated as
\begin{equation}
{\boldsymbol{\hat{\mu}}}=\frac{1}{N-2}\sum_{k=1}^{N-2}{\mathbf{t}\left( k,k+2 \right)}  \label{eq42}
\end{equation}
and
\begin{equation}
\mathbf{\hat{\Phi}}=\frac{1}{N-3}\sum_{k=1}^{N-2}{{\mathbf{I}}_{2D} \cdot \left\{ \left[ \mathbf{t}\left( k,k+2 \right) -{\boldsymbol{\hat{\mu} }} \right] \circ \left[ \mathbf{t}\left( k,k+2 \right) -{\boldsymbol{\hat{\mu} }} \right] \right\}} \label{eq43}
\end{equation}
respectively. Then, we construct a test statistic as follows
\begin{equation}
%\mathcal{T}=\mathbf{\hat{q} '}^T\mathbf{\hat{\Phi}'}^{-1}\mathbf{\hat{q} '}
{\cal T} =  {\bf{q }}\,^T {\bf{\hat{\Phi }}}\,^{ - 1} {\bf{q }}. \label{eq44}
\end{equation}

Theoretically, the test statistic ${\cal T}={\bf{p}}^T {\bf{p}}$ asymptotically follows a chi-square distribution with $2D$ degrees of freedom, i.e., ${\cal T} \to \chi _{2D}^2$. However, since ${\boldsymbol{\mu }} \ne 0$ and is unknown, \eqref{eq42} suffers from a certain error between ${\boldsymbol{\mu }}$ and ${\boldsymbol{\hat{\mu} }}$ for a limited observation period even though ${\boldsymbol{\hat{\mu} }}$ is an asymptotically unbiased estimator, which impacts the distribution of ${\cal T}$ in practice. Let ${\boldsymbol{\mu}} = {\boldsymbol{\hat{\mu} }}   + \Delta {\boldsymbol{\mu}} $ with a small deviation $\Delta {\boldsymbol{\mu}}$. Then, ${\cal T}$ approximately follows a non-central chi-square distribution with $2D$ degrees of freedom and its CDF is given by \cite{probability_handbook}
\begin{equation}
\Pr \left( { {\cal T} < \lambda } \right) = 1 - Q_{D} \left( {\left\| {\Delta {\boldsymbol{\mu}} } \right\|_F,\sqrt \lambda } \right),  \lambda \ge 0  \label{eq45}
\end{equation}
where $Q \left( \cdot \right)$ is the generalized ($m$-order) Marcum $Q$-function defined as
\begin{equation}
Q_m \left( {\alpha ,\beta } \right) = \frac{1}{{\alpha ^{m - 1} }}\int_\beta ^\infty  {t^m \exp \left( { - \frac{{t^2  + \alpha ^2 }}{2}} \right)} J_m \left( {\alpha t} \right){\rm{d}}t  \label{eq46}
\end{equation}
with the modified Bessel function $J_m \left( \cdot \right)$ of order $m$ \cite{probability_handbook}.

\begin{figure}
  \centering
  \includegraphics[width=0.5\textwidth]{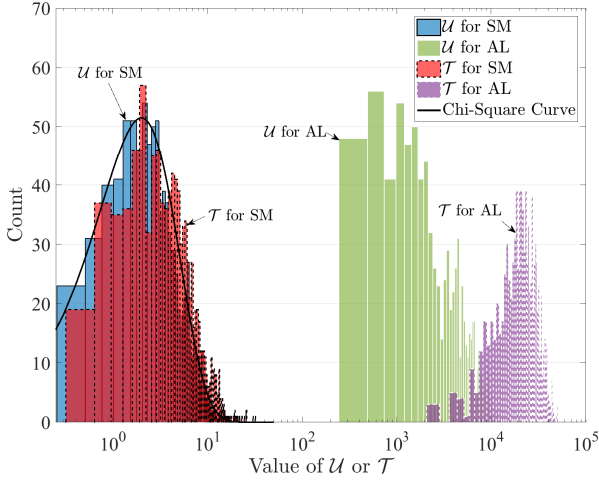}\\
  \caption{Histogram of the test statistics, where $N=256$, $\Omega = \{ (1,2), (2,1)\}$, $N_b=100$ and SNR $=10$ dB. In addition, $G=1$ for HT-based algorithm. The simulation was run for 1000 trials.}\label{His}
\end{figure}

For {\bf AL-SFBC}, it is complicated to calculate $\cal T$ due to the absolute value operations. However, we can still conclude that we have ${\cal T} > {\cal U}$ in the high-SNR regime or under a large $N_b$ as discussed next.

\emph{Proposition 3:} We define that ${\bf A} \ge {\bf B}$ if any element of $\bf A$, denoted by ${\bf A}_{ij}$, is greater than or equal to the element at the corresponding location of $\bf B$, denoted by ${\bf B}_{ij}$, i.e., ${\bf A}_{ij} \ge {\bf B}_{ij}$. Then, we have $\mathbf{\Psi  } \ge  \mathbf{\Phi }$.

\emph{Proof:} See Appendix D.

From the \emph{proof} of \emph{Proposition 3}, $\mathbf{\Psi } \ge{{\mathbf{I}}_{2D}}{\boldsymbol{\mu}}$. In the high-SNR regime or under a large $N_b$,  $\mathbf{t }(2j-1,2j) \gg  \mathbf{ t }(k,k+2) \approx {\boldsymbol{\mu}}$, and hence, we can regard ${\boldsymbol{\mu}}$ as an approximate zero vector compared with a large $\mathbf{t }(2j-1,2j)$. Then, we have $\mathbf{q } \ge \mathbf{ v }_i$ since each term at the right hand side of \eqref{eq41} is the absolute value of the corresponding term of \eqref{eq18}. Therefore, we have ${\cal T} > {\cal U}$ in the high-SNR regime or under a large $N_b$. Fig. \ref{His} shows that $\cal T$ is more distinguishable than $\cal U$ between SM and AL signals. The histograms of $\cal T$ and $\cal U$ almost overlap for SM signals but differ significantly for AL signals.

The hypothesis test approach is not suitable for making the decision on $\cal T$ due to the unknown ${\Delta {\boldsymbol{\mu}} }$. After calculating $\cal T$, the discriminating problem can be considered as a two-class classification problem. Given that SVM is a powerful classification algorithm, since the optimality criterion is convex and it is robust over different training samples \cite{bishop2006pattern}, we employ the SVM algorithm to make the decision. The SVM constructs an optimal hyperplane in a high-dimensional space which can be used for classification based on the test statistic $\cal T$. The hyperplane has the largest distance to the nearest training data point of any class. Generally, the SVM processing has two main steps: training and testing. The first step is to determine the optimal hyperplane separating SM and AL signals by using the training data obtained from known sources. In this paper, the kernel  and soft margin parameter are set to linear kernel and 1, respectively, since $\cal T$ is strongly distinguishable and linearly separable. In addition, the SVM should be retrained when changing the number of receive antennas because the degree of freedom of the distribution for $\cal T$ changes. In the second step, the test data is compared with the trained hyperplane and then classified accordingly. 

 For clarity, the main steps of the proposed algorithm are summarized below.
  \begin{algorithm}
 \caption{}
 \begin{algorithmic}[1]
 \renewcommand{\algorithmicrequire}{\textbf{Input: }}
 \renewcommand{\algorithmicensure}{\textbf{Output:}}
 \REQUIRE The observed sequence $\bf{y}$ and trained SVM.
 \ENSURE  SFBC.
 \STATE{ Construct the stacked vectors $\mathbf{{t}}\left( 2j-1,2j \right)$, $j = 1, 2, \cdots , N/2$, and $\mathbf{{t}}\left( k,k+2 \right)$, $k = 1,2, \cdots ,N-1$, using \eqref{eq39}.}
 \STATE{ Compute the mean vector ${\boldsymbol{\hat{\mu}}}$ using \eqref{eq42} and then $\mathbf{q}$ using \eqref{eq41}.} 
 \STATE{ Compute the covariance matrix $\mathbf{\hat{\Phi}}$ using \eqref{eq43}. } 
 \STATE{ Construct the test statistic ${\cal T}$ using \eqref{eq44}.}
 \STATE {The SVM makes the decision.}
 \RETURN {SFBC.}
 \end{algorithmic}
 \end{algorithm}

\section{Simulation Results}

\subsection{Simulation Setup}

Monte Carlo simulations are conducted to evaluate the performance of the proposed algorithms. Unless otherwise stated, we consider a MIMO-OFDM system with $N_r=2$ receive antennas, the set of receive antenna pairs $\Omega = \{ (1,2), (2,1)\}$, $N=512$ sub-carriers, cyclic prefix length $\nu = 10$, and QPSK modulation. For the HT-based algorithm, the default value of $G$ was set to 8. In addition, we assume two transmit antennas transmitting both SM and AL-SFBC signals. The channel is assumed to be frequency-selective and consists of $L_h=4$ statistically independent taps with an exponential power delay profile\cite{Blind_MIMO_OFDM_SM_AL}, $\sigma _{\tau}^2 = {e^{ - {{\tau} / 5}}}$, where ${\tau} = 0, \cdots ,{L_h} - 1$. The probability of false alarm $\Pr _f$ was set to $10^{-3}$ and the number of observed OFDM symbols $N_b$ was 20. The SNR is defined as $10\log _{10} \left( {P/\sigma_n ^2 } \right)$ with $P = 1$ and $\sigma _n^2$ being the total transmit power and the AWGN variance, respectively. The probability of correct identification ${\Pr} = 0.5{\Pr}\left( {{\rm{SM}}\left| {{\rm{SM}}} \right.} \right) + 0.5{\Pr}\left( {{\rm{AL}}\left| {{\rm{AL}}} \right.} \right)$, was used as a performance measure. Simulation of each SFBC type was run for 1000 trials. For the training of the SVM, we set the system parameters as we mentioned previously and generate the datasets from 0 dB to 15 dB, where each SNR repeats 50 Monte Carlo trials for both codes.

\subsection{Performance Evaluation}

\begin{figure}
\centering
\includegraphics[width=0.5\textwidth]{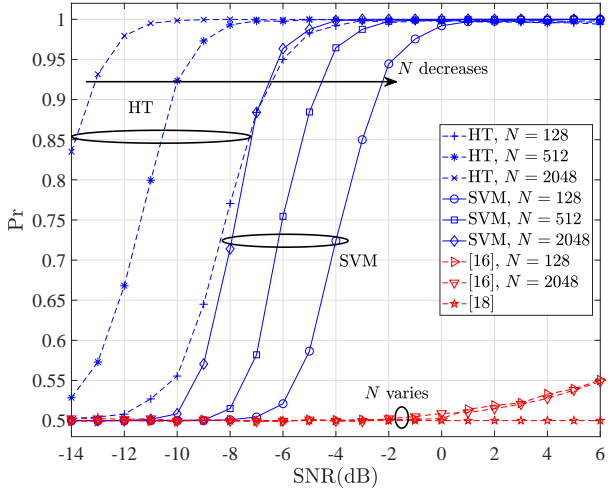}
\caption{Performance comparison of the proposed algorithms and the algorithms in \cite{blind_SFBC,My_paper_TVT} for different $N$ based on the average probability of correct identification $\Pr$ under the same conditions. }\label{fig3}
\end{figure}

\begin{figure}
\centering
\includegraphics[width=0.5\textwidth]{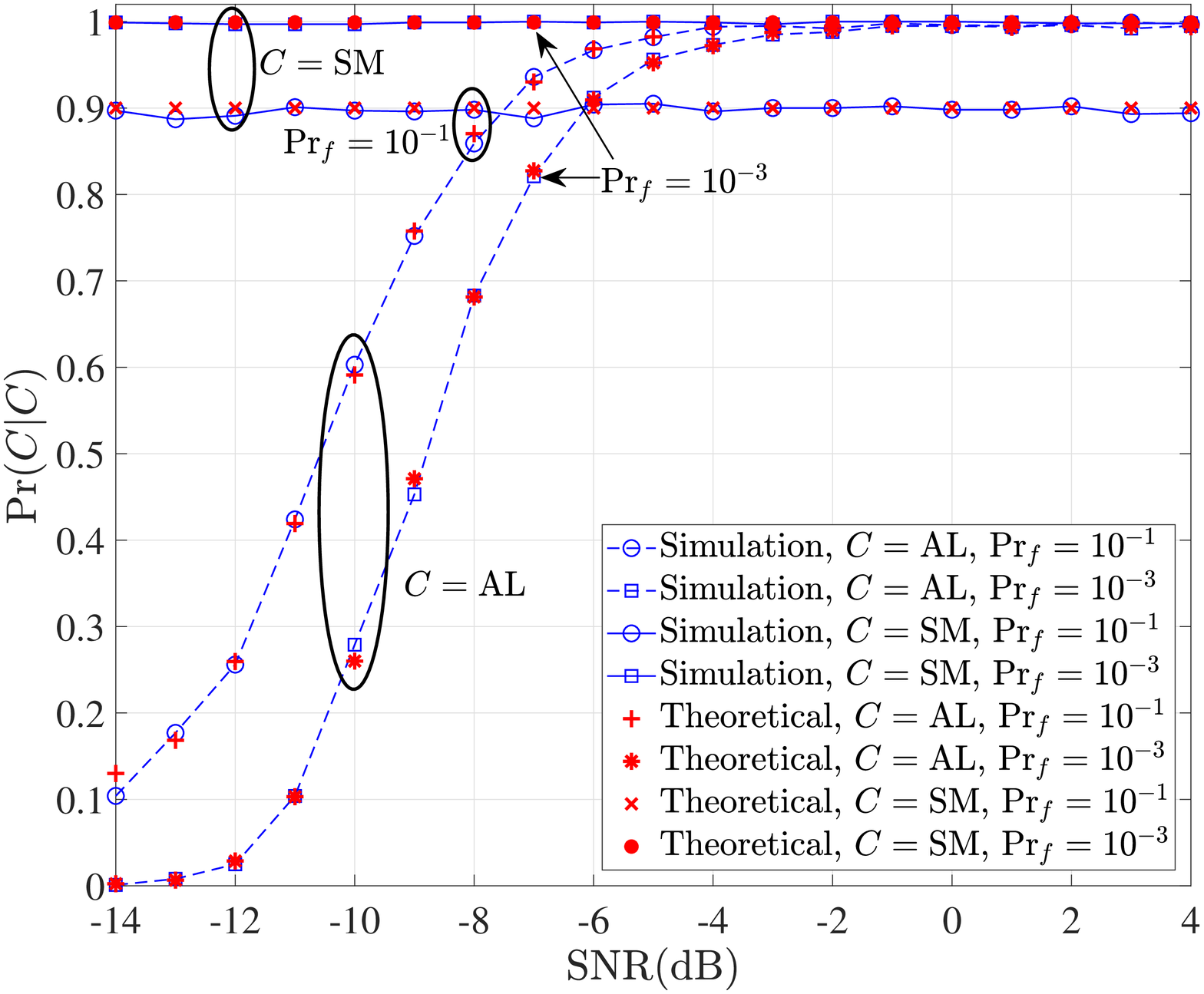}
 \caption{Simulation and theoretical results for various false alarm probabilities, ${P_f}$, on the average probability of correct identification $\Pr(C|C)$ for the HT-based algorithm. }\label{fig4}
\end{figure}

Fig. \ref{fig3} shows the performance of the proposed HT- and SVM-based algorithms in comparison with those in \cite{blind_SFBC} and \cite{My_paper_TVT} for different numbers of OFDM sub-carriers under the same conditions. The set of time lags $\Upsilon$ in \cite{blind_SFBC} was set to $\{ 0,1,2, 3,4,5,6 \}$ with cardinality ${\rm card}\left( \Upsilon  \right) = 7$. The simulation results demonstrate that our proposed algorithms significantly outperform the algorithms in \cite{blind_SFBC} and \cite{My_paper_TVT}, while the probability of correctly identifying SFBC signals employing the algorithm in \cite{blind_SFBC} is independent of $N$. This is because the convergence of a normalized random variable depends on the number of OFDM sub-carriers $N$ as shown in \eqref{eq18} and \eqref{eq41}, and the cross-correlation function in \cite{blind_SFBC} does not depend on $N$. Moreover, the algorithm in \cite{blind_SFBC} requires a larger number of OFDM symbols or receive antennas to achieve the same performance. The algorithm in \cite{My_paper_TVT} fails to identify the SFBC signals when the number of receive antennas is equal to the number of transmit antennas. 

From a practical point of view, we analyze the computational complexity between the proposed algorithms and the algorithms in \cite{blind_SFBC,My_paper_TVT}, as summarized in Table \ref{table2}. Based on the number of floating point operations (flops) definitions in \cite{matrix_computations}, the main computational complexity of the HT- and SVM-based algorithms is given by $8{N_b} {N }  {D} $. Here, the number of flops for a complex multiplication and addition are 6 and 2, respectively. Meanwhile, the main computational complexities of the algorithms in \cite{blind_SFBC,My_paper_TVT} are given by $4{N_b}\left( {N + \nu } \right) {D} \left( {{\rm card}\left( \Upsilon  \right) + 1} \right)$ and $0.75N\left( {64N_r^3 + 32N_r^2{N_b}} \right)$, respectively. In the previous case, i.e., $N=512$, $\nu = 10$, $N_b = 20$, $D = 2$, the proposed algorithms require approximately 0.2 Mega-flops. Employing a low-power TMS320C6742 processor with 1.2 Giga-flops \cite{DSP}, the proposed algorithms require an execution time of 140 $\mu $s, while the LTE standard requires about 1.43 ms for transmitting 20 OFDM symbols with one block duration of 71.4 $\mu $s \cite{sesia2009lte}. We can also see that the proposed algorithms have lower computational complexity although they achieve significantly better performance as shown in Fig. \ref{fig3}.

\begin{table}[htbp]
\centering
\caption{FLOPS comparison among the proposed algorithms and those in \cite{blind_SFBC} and \cite{My_paper_TVT} for $N=512,N_{b}=20,N_r=2$}
\label{table2}
\begin{tabular}{@{}ccc@{}}
\toprule
Algorithm                  & Main computational cost                                           & Number of flops                \\ \midrule
HT                             & $8{N_b} {N }  {D} $     &   163,840                      \\
SVM                           &  $8{N_b} {N }  {D} $    &    163,840        \\           
\cite{blind_SFBC}        & $4{N_b}\left( {N + \nu } \right) {D} \left( {{\rm card} (\Upsilon)   + 1} \right)$     & 668,160                    \\
 \cite{My_paper_TVT}     & $0.75N\left( {64N_r^3 + 32N_r^2{N_b}} \right)$                         & 1,179,684                      \\ \bottomrule
\end{tabular}
\end{table}

Fig. \ref{fig4} shows the theoretical and simulation results of the HT-based algorithm for the probability of correctly identifying SM and AL signals for various probabilities of false alarm, $\Pr _f$. Here, we used the simplest set of receive antenna pairs $\Omega = \{ (1,2)\}$, $N_b=100$ and $G = 1$. In general, the theoretical expressions and the simulation results are in good agreement. The SM identification performance decreases with an increase in $\Pr _f$, as ${\Pr}({\rm{SM|SM}}) = 1 - {\Pr _f}$. On the other hand, the AL identification performance improves as $\Pr _f$ increases. This results from the reduction in the threshold value $\eta$.

\subsection{Identification of 3-antenna SFBCs}

\begin{figure}
\centering
\includegraphics[width=0.5\textwidth]{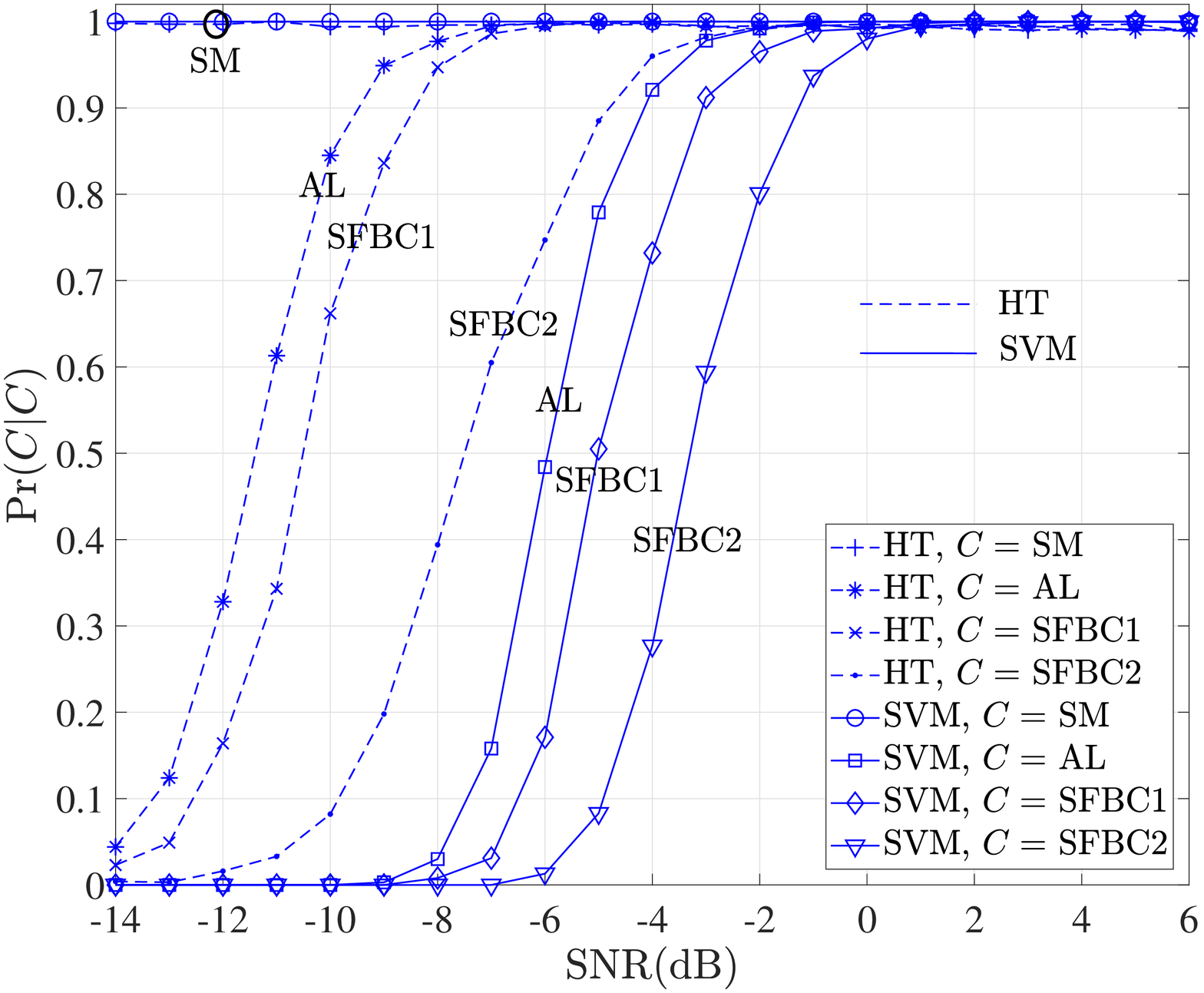}
 \caption{ Simulation for different SFBCs on the average probability of correct identification $\Pr(C|C)$. }\label{SFBC3}
\end{figure}

Fig. \ref{SFBC3} shows the results of the proposed algorithms for the probability of correctly identifying SM, AL, SFBC1 and SFBC2 signals using the decision tree identification. We can see that the performance of identifying AL signals is better than that of 3-antenna SFBC signals. 

%{\color{blue}{This is because ${R}_{\rm SFBC1}^{\left( i_1, i_2 \right) }\left( k,k+4 \right) $, ${R}_{\rm SFBC2}^{\left( i_1, i_2 \right) }\left( k,k+2 \right)$ and ${{R}_{\rm AL}^{\left( i_1, i_2 \right) }}\left( {k,k + 1} \right)$ are different in \eqref{eq10}-\eqref{eq12}. Statistically, we have
%\begin{equation}
%{{R}_{\rm AL}^{\left( i_1, i_2 \right) }}\left( {k,k + 1} \right) > {R}_{\rm SFBC1}^{\left( i_1, i_2 \right) }\left( k,k+4 \right)  > {R}_{\rm SFBC2}^{\left( i_1, i_2 \right) }\left( k,k+2 \right)
%\end{equation} 
%since the latter has more independent and identically distributed random terms in their expressions and converges to zero in probability according to the law of large numbers.}}

\begin{figure}
 \centering
 \includegraphics[width=0.5\textwidth]{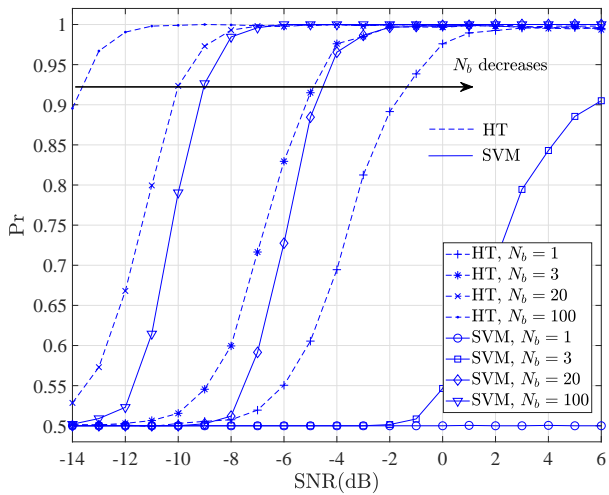}
  \caption{Effect of the number of observed OFDM symbols, ${N_b}$, on the average probability of correct identification $\Pr$. }\label{fig5}
\end{figure}

\subsection{Effect of the Number of Processed OFDM Symbols}

Fig. \ref{fig5} illustrates the performance of the proposed algorithms for different numbers of OFDM symbols. We can see that the performance of these two algorithms improves with the number of OFDM symbols since $\epsilon$ vanishes. It can also be seen that the HT-based algorithm can identify SM and AL signals even using one OFDM symbol and the SVM-based algorithm only requires  three OFDM symbols owing to its effective utilization of the redundant information among the OFDM sub-carriers. In such case, the HT-based algorithm requires approximately 0.01 Mega-flops while  the SVM-based algorithm requires approximately 0.03 Mega-flops. This indicates that our proposed algorithms can implement real-time processing after receiving a very small number of OFDM symbols and satisfy the requirement of delay-sensitive services, which are important in next generation networks.

\subsection{Effect of the Number of Receive Antennas}

\begin{figure}
\centering
\includegraphics[width=0.5\textwidth]{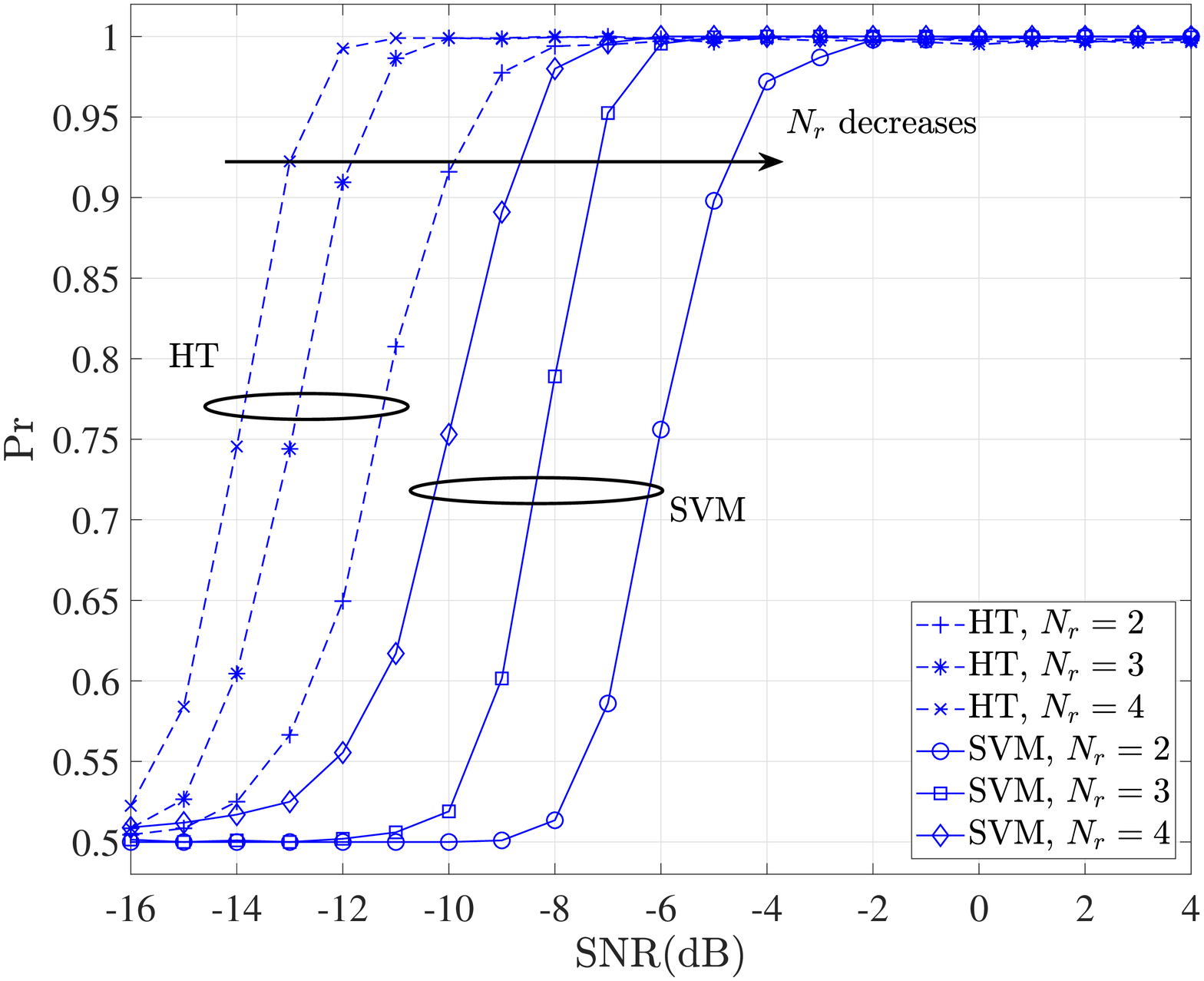}
  \caption{Effect of the number of receive antennas, ${N_r}$, on the probability of correct identification $\Pr$. }\label{fig6}
\end{figure}

\begin{figure}
 \centering
 \includegraphics[width=0.5\textwidth]{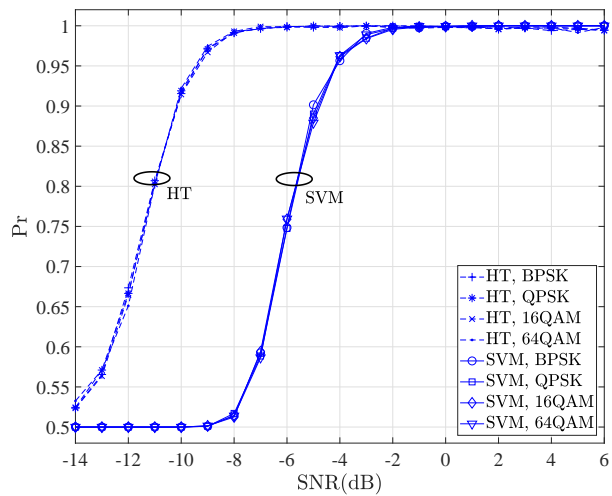}
  \caption{Effect of the modulation type on the average probability of correct identification $\Pr$. }\label{fig7}
\end{figure}

Fig. \ref{fig6} shows that the probability of correct identification improves with the number of receive antennas (the number of elements in $\Omega$ is maximized here). In fact,  $\cal U$ and $\cal T$ increase significantly with $N_r$ when the received signals are estimated as AL since the sum of the constant terms on the right hand of \eqref{eq38} increases, which in turn results in a higher ${\Pr}({\rm{AL|AL}})$. 

\subsection{Effect of the Modulation Type}

Fig. \ref{fig7} illustrates the effect of the modulation type on the identification performance. The performance does not depend on the modulation type. This can be explained by the fact that the cross-correlation function described in \eqref{eq8} applies to both $M$-QAM and $M$-PSK modulations regardless of the modulation order. This feature provides the designer with the ability to implement the modulation classifier either before or after the proposed SFBC identification algorithms.

\subsection{Effect of the Timing Offset}

\begin{figure}
 \centering
 \includegraphics[width=0.5\textwidth]{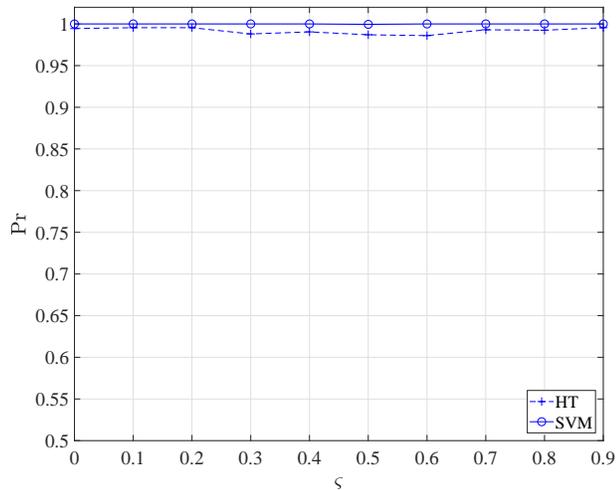}
  \caption{Effect of the sampling clock offset, $\varsigma$, on the average probability of correct identification $\Pr$. }\label{fig8}
\end{figure}

\begin{figure}
\centering
\includegraphics[width=0.5\textwidth]{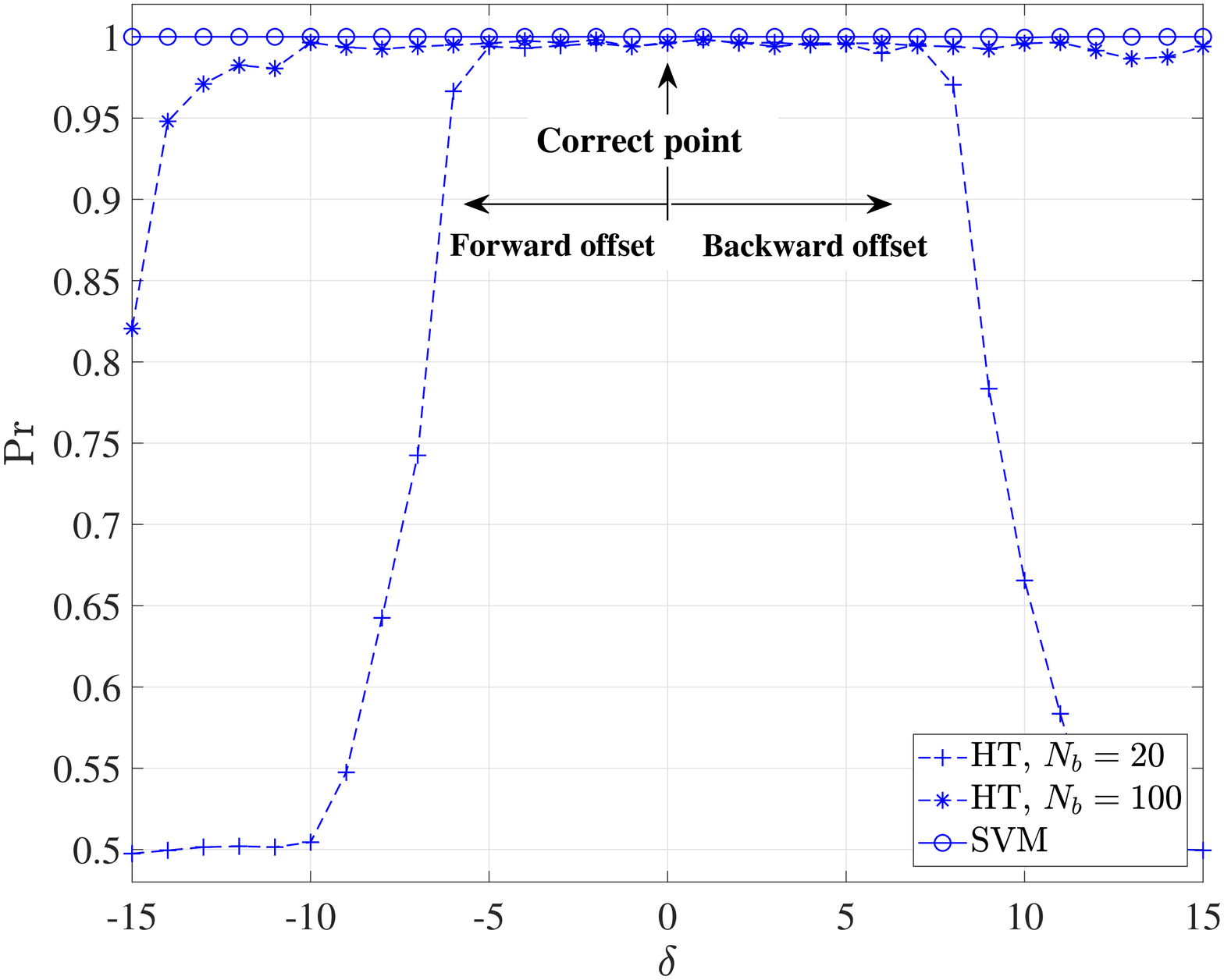}
  \caption{Effect of STO, $\delta$, on the average probability of correct identification $\Pr$. }\label{fig9}
\end{figure}

The simulation results presented so far have been under perfect timing synchronization. Now, we evaluate the performance of the proposed algorithms under timing offsets. The timing offset has two components, namely, sampling clock offset and symbol timing offset (STO). The effect of the sampling clock offset can be modeled as a two-path channel $[1-\varsigma, \varsigma ]$ \cite{time_offset}, where $0 \le \varsigma < 1 $ is the normalized sampling clock offset when the whole sampling period is one. The STO is modeled as in \cite{MIMO-OFDM_with_MATLAB}, which depends on the location of the estimated FFT window starting point of OFDM symbols, denoted by $\delta $. Figs. \ref{fig8} and \ref{fig9} show the performance of the proposed algorithms for different sampling clock offsets and STOs, respectively. The SNR was set to 6 dB in these figures. We can see that the proposed algorithms are essentially not affected by the sampling clock offset while the HT-based algorithm fails under a large STO. For the HT-based algorithm, the STO of $\delta $ in time domain incurs the phase rotation of $2\pi k\delta /N$ in the frequency domain, which is proportional to the OFDM sub-carrier index $k$ as well as to the STO $\delta$. After these phase rotations, the values of $\hat{R}^{\left( i_1,i_2 \right)}\left( k_1,k_2 \right)$ are distributed uniformly on the complex plane and have zero mean which results in the first term on the right hand of  \eqref{eq28} approaching zero. As for the SVM-based algorithm, the effect of the phase rotations is eliminated by the absolute value operations. 
%{\color{blue}{In addition, Fig. \ref{fig8} also indicates that the SVM-based algorithm performs better than the HT-based algorithm in the high-SNR regime, since $\cal T$ is more distinguishable than $\cal U$ between SM and AL signals in the high-SNR regime as we previously described.}} 

\subsection{Effect of the Frequency Offset}

\begin{figure}
 \centering
 \includegraphics[width=0.5\textwidth]{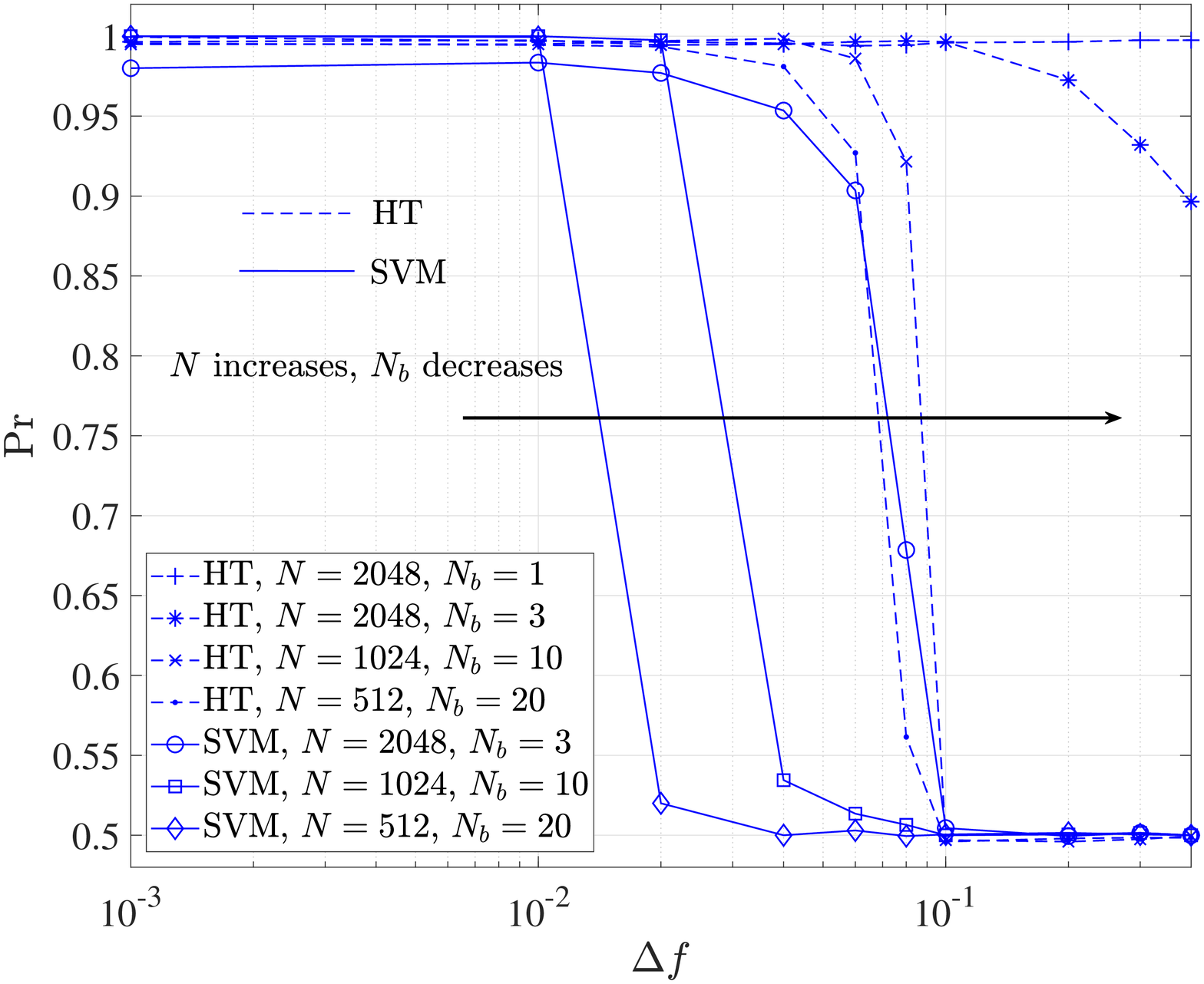}
  \caption{Effect of the frequency offset, $\Delta f$, on the average probability of correct identification $\Pr$. }\label{fig10}
\end{figure}

\begin{figure}
\centering
\includegraphics[width=0.5\textwidth]{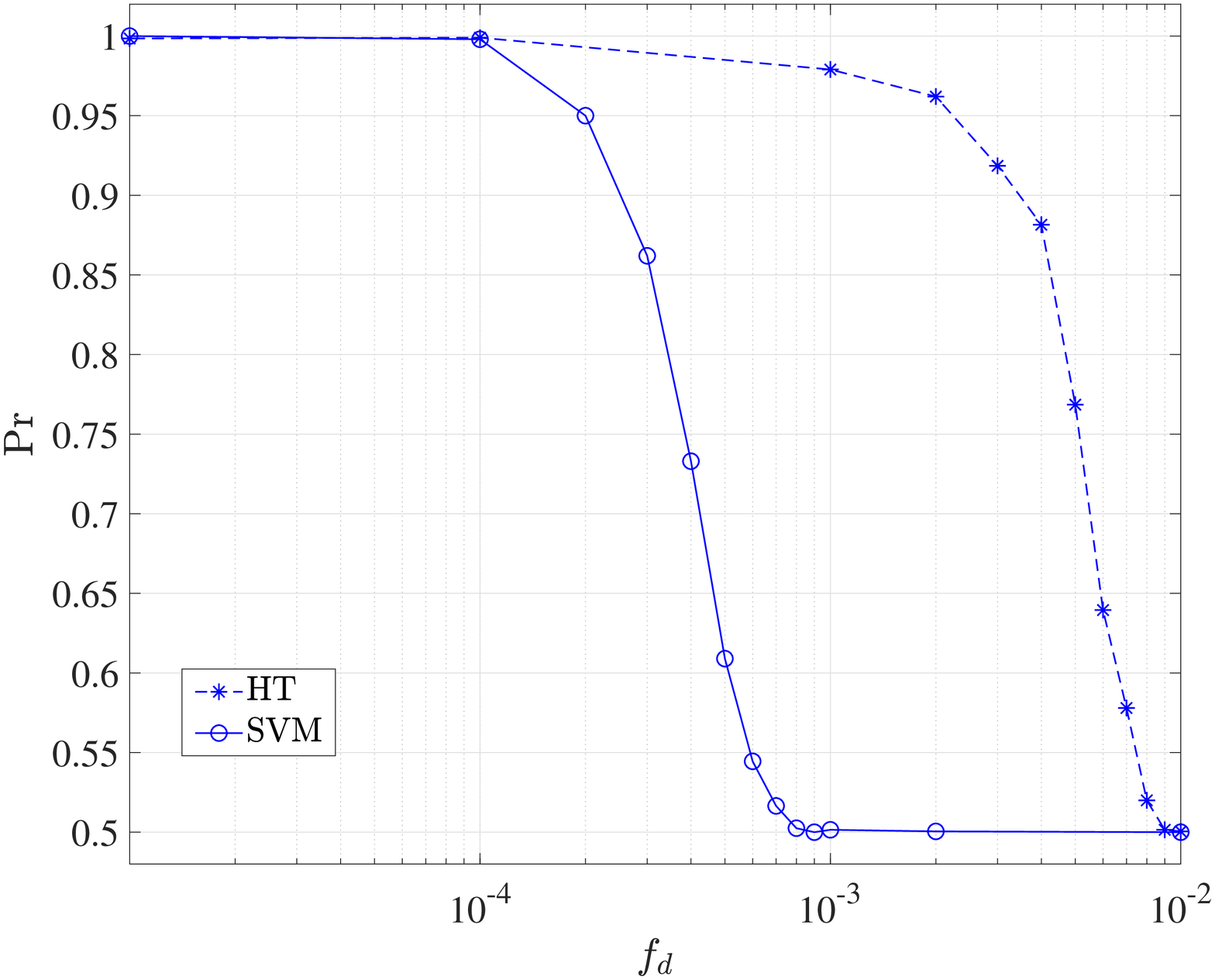}
  \caption{The effect of the Doppler frequency, $f_d$, on the average probability of correct identification $\Pr$. }\label{fig11}
\end{figure}

Fig. \ref{fig10} illustrates the effect of the frequency offset normalized to the OFDM sub-carrier spacing, $\Delta f$, on the performance of the proposed algorithms   at SNR $=$ 6 dB and for different values of $N$ and $N_b$. The frequency offset undoubtedly destroys the orthogonality of AL-SFBC signals \cite{ICI_affects_OSTBC} and degrades the performance. It is worth noting that a smaller number of OFDM symbols is required to achieve a good performance for a large number of OFDM sub-carriers, which results in a lower sensitivity to the frequency offset. The results in Fig. \ref{fig10} show a generally good robustness for $\Delta f < 10^{-2}$ for the proposed algorithms. Further, the HT-based algorithm has a good robustness for  $\Delta f < 10^{-1}$ when $N=2048$ and $N_b \le 3$. Furthermore, we can use a blind frequency offset compensation technique \cite{blind_freq_offset} by utilizing the kurtosis-type criterion before OFDM domodulation to reduce the effect of the frequency offset.

\subsection{Effect of the Doppler Frequency}
The previous analysis assumed static channels over the observation period. The typical parameters of the LTE standard with the channel bandwith of 10 MHz ($N=512$) and samping rate of 15.36 MHz are assumed here to evaluate the impact of the Doppler frequency on the performance of the proposed algorithms. Fig. \ref{fig11} shows the average probability of correct identification versus the maximum Dopper frequency normalized to the sampling rate, $f_d$, at SNR $=$ 6 dB. The results for the HT- and the SVM-based algorithms show a good robustness for $f_d < 10^{-3}$ and $f_d < 10^{-4}$, i.e., 15360 Hz and 1536 Hz, respectively. In other words, the HT-based algorithm is robust to the highest Doppler shift for mobile speeds up to 8290 km/h (5150 MPH), while the SVM-based algorithm is up to 829 km/h (515 MPH) for an LTE system at a carrier frequency of 2 GHz. 
%In addition, the proposed algorithms also have relatively good robustness against a small non-Gaussian noise \cite{Non_Gaussian}.

%{\color{blue}{
%\subsection{Effect of the Impulsive Noise}
%Impulsive noise can be used as the model of the interference from other transmitter-receiver pairs, namely co-channel interference. Fig. \ref{fig12} shows the effect of the impulsive noise on the proposed algorithm. Here the impulsive noise is modeled as the Gaussian mixture noise with the probability density function (PDF) given by \cite{Non_Gaussian}
%\begin{equation}
%p(t) = (1- \varepsilon ){\cal N}(0, \sigma^2) + \varepsilon{\cal N}(0, \eta \sigma^2)
%\end{equation}
%where $\varepsilon$ is the probability of impulses in noise and ${\cal N}(0, \sigma^2)$ and ${\cal N}(0, \eta \sigma^2)$ denote zero mean Gaussian PDFs with variances $\sigma^2$ and $\eta \sigma^2$, respectively. The results indicate that the proposed algorithm has relatively good robustness against the small impulsive noise.
%}}
%\begin{figure}
%  \centering
%  \includegraphics[width=0.5\textwidth]{Impulsive.eps}\\
%  \caption{Effect of the impulsive noise on the average probability of correct identification $\rm{Pr}$.}\label{fig12}
%\end{figure}

\section{Conclusion}

Based on the CLT, we proposed two novel algorithms, namely HT-based and SVM-based algorithms, to blindly identify SFBC signals over frequency-selective fading channels. The two algorithms use the cross-correlation function of the received signals from antenna pairs at consecutive OFDM sub-carriers to exploit the space-domain redundancy. The HT-based algorithm utilizes the frequency-domain redundancy by constructing a chi-square test statistic that is used to make the decision. In addition, the theoretical expressions of  the probability of correctly identifying the SM and AL-SFBC signals for the HT-based algorithm are derived. The SVM-based algorithm uses a strongly-distinguishable non-central chi-square statistical feature to exploit the frequency-domain redundancy and employs a trained SVM to make the decision. The proposed algorithms can improve the identification performance since they exploit additional redundancy in the signal structure. Furthermore, they have a low computational complexity and do not require \textit{apriori} knowledge about the channel coefficients, modulation type or noise power. Moreover, the SVM-based algorithm does not require timing synchronization. Simulation results demonstrated that a good identification performance is achieved under frequency-selective fading with a short observation period. Furthermore, the HT-based algorithm has a good robustness to small STOs, and the two proposed algorithms show a relatively good robustness to frequency offsets and Doppler effects. Based on the features of the two proposed algorithms, we conclude that the HT-based algorithm can be used in a relatively low SNR-regime with timing synchronization, while the SVM-based algorithm can be used in the so-called ``totally-blind'' applications, such as military communications. In addition, blind identification in multi-user cases is still unexplored, and represents a direction for future work.

\appendices
\section{Proof of Proposition 1}
From \eqref{eq7} and \eqref{eq14}, the mean of ${\epsilon} ^{(1,2)} (k_1, k_2) $ is given by
\begin{align}
{\rm{E}}\left[ {{\epsilon} ^{(1,2)} \left( {k_1 ,k_2 } \right)} \right] =& \mathop {\lim }\limits_{N_b  \to \infty } \frac{1}{{N_b }}\sum\limits_{n = 1}^{N_b } {({\bf{H}}_{k_1 }^{(1)} {\bf{s}}_{k_1 } (n){\bf{H}}_{k_2 }^{(2)} {\bf{s}}_{k_2 } (n)}  \notag \\
& + w_{k_2 }^{(2)} (n){\bf{H}}_{k_1 }^{(1)} {\bf{s}}_{k_1 } (n)  \notag \\
&+ w_{k_1 }^{(1)} (n){\bf{H}}_{k_2 }^{(2)} {\bf{s}}_{k_2 } (n) \notag \\
&+ w_{k_1 }^{(1)} (n)w_{k_2 }^{(2)} (n))  = 0.
\end{align}
The covariance matrix is a diagonal matrix since the elements ${\boldsymbol{\epsilon }}(k_1,k_2)$ are independent of each other. For convenience, we simplify ${{\epsilon} ^{(1,2)} \left( {k_1 ,k_2 } \right)}$ as ${{\epsilon} \left( {k_1 ,k_2 } \right)}$. Suppose that the covariance matrix is a diagonal matrix and given by ${\bf{\Psi }} = {\rm{diag}}\left( {\sigma _{\epsilon _1 }^2 ,\sigma _{\epsilon _2 }^2 } \right)$. According to our assumptions, for a large $N_b$, we have the derivation expressed as in \eqref{eq49}
\setcounter{TempEqCnt}{\value{equation}}
\setcounter{equation}{48}
\begin{figure*}[ht]
  \normalsize
%  \vspace*{4pt}
\begin{align} \label{eq49}
\sigma _{\epsilon _1 }^2  =& \  {\rm{E}}\left[ {\left( {\Re \left\{ {\epsilon \left( {k_1 ,k_2 } \right)} \right\}} \right)^2 } \right] - {\rm{E}}\left[ {\Re \left\{ {\epsilon \left( {k_1 ,k_2 } \right)} \right\}} \right]^2  \notag \\ 
=& \   \Re \left\{ {\frac{1}{2}{\rm{E}}\left[ {\left( {\Re \left\{ {\epsilon \left( {k_1 ,k_2 } \right)} \right\}} \right)^2  + \left( {\Im \left\{ {\epsilon \left( {k_1 ,k_2 } \right)} \right\}} \right)^2  + \left( {\Re \left\{ {\epsilon \left( {k_1 ,k_2 } \right)} \right\}} \right)^2  - \left( {\Im \left\{ {\epsilon \left( {k_1 ,k_2 } \right)} \right\}} \right)^2 } \right]} \right\}  \notag \\
 =& \  \Re \left\{ {\frac{1}{2}{\rm{E}}\left[ {\epsilon \left( {k_1 ,k_2 } \right)\epsilon ^ *  \left( {k_1 ,k_2 } \right) + \epsilon \left( {k_1 ,k_2 } \right)\epsilon \left( {k_1 ,k_2 } \right)} \right]} \right\} = \Re \left\{ {\frac{1}{2}{\rm{E}}\left[ {\epsilon \left( {k_1 ,k_2 } \right)\epsilon ^ *  \left( {k_1 ,k_2 } \right)} \right]} \right\} \notag \\
 =& \  \Re \{ \frac{1}{2}(\mathop {\lim }\limits_{N_b  \to \infty } \frac{1}{{N_b }}\sum\limits_{n = 1}^{N_b } {({\bf{H}}_{k_1 }^{(1)} {\bf{s}}_{k_1 } (n){\bf{H}}_{k_2 }^{(2)} {\bf{s}}_{k_2 } (n) + w_{k_2 }^{(2)} (n){\bf{H}}_{k_1 }^{(1)} {\bf{s}}_{k_1 } (n)}  + w_{k_1 }^{(1)} (n){\bf{H}}_{k_2 }^{(2)} {\bf{s}}_{k_2 } (n) \notag \\
& + w_{k_1 }^{(1)} (n)w_{k_2 }^{(2)} (n))) \cdot (\mathop {\lim }\limits_{N_b  \to \infty } \frac{1}{{N_b }}\sum\limits_{n = 1}^{N_b } {({\bf{H}}_{k_1 }^{(1) * } {\bf{s}}_{k_1 }^ *  (n){\bf{H}}_{k_2 }^{(2) * } {\bf{s}}_{k_2 }^ *  (n) + w_{k_2 }^{(2) * } (n){\bf{H}}_{k_1 }^{(1) * } {\bf{s}}_{k_1 }^ *  (n)}  \notag \\
& + w_{k_1 }^{(1) * } (n){\bf{H}}_{k_2 }^{(2) * } {\bf{s}}_{k_2 }^ *  (n) + w_{k_1 }^{(1) * } (n)w_{k_2 }^{(2) * } (n)))\}  \notag \\
 =& \  \frac{1}{{2N_b }}({\bf{H}}_{k_1 }^{(1)} {\rm{E}}\left[ {{\bf{s}}_{k_1 } (n){\bf{s}}_{k_1 }^H (n)} \right] {\bf{H}}_{k_1 }^{(1)H}{\bf{H}}_{k_2 }^{(2)}{\rm{E}}\left[ {{\bf{s}}_{k_2 } (n){\bf{s}}_{k_2 }^H (n)} \right] {\bf{H}}_{k_2 }^{(2)H}  +  {\rm{E}}\left[ {w_{k_2 }^{(2)} (n)w_{k_2 }^{(2) * } (n)} \right]{\bf{H}}_{k_1 }^{(1)}{\rm{E}}\left[ {{\bf{s}}_{k_1 } (n){\bf{s}}_{k_1 }^H (n)} \right]{\bf{H}}_{k_1 }^{(1)H}  \notag \\
& +  {\rm{E}}\left[ {w_{k_1 }^{(1)} (n)w_{k_1 }^{(1) * } (n)} \right]{\bf{H}}_{k_2 }^{(2)}{\rm{E}}\left[ {{\bf{s}}_{k_2 } (n){\bf{s}}_{k_2 }^H (n)} \right]{\bf{H}}_{k_2 }^{(2)H}  + {\rm{E}}\left[ {w_{k_1 }^{(1)} (n)w_{k_1 }^{(1) * } (n)} \right]{\rm{E}}\left[ {w_{k_2 }^{(2)} (n)w_{k_2 }^{(2) * } (n)} \right])  \notag \\
 = & \  \frac{{\sigma _s^4 }}{{2N_b }}\left\| {{\bf{H}}_{k_1 }^{(1)} } \right\|_F^2 \left\| {{\bf{H}}_{k_2 }^{(2)} } \right\|_F^2  + \frac{{\sigma _s^2 \sigma _w^2 }}{{2N_b }}\left( {\left\| {{\bf{H}}_{k_1 }^{(1)} } \right\|_F^2  + \left\| {{\bf{H}}_{k_2 }^{(2)} } \right\|_F^2 } \right) + \frac{{\sigma _w^4 }}{{2N_b }}.
\end{align}
\end{figure*}

Similarly, 
\begin{align}
\sigma _{\epsilon _2 }^2  &= {\rm{E}}\left[ {\left( {\Im \left\{ {\epsilon \left( {k_1 ,k_2 } \right)} \right\}} \right)^2 } \right]  \notag \\
&= \Re \left\{ {\frac{1}{2}{\rm{E}}\left[ {\epsilon \left( {k_1 ,k_2 } \right)\epsilon ^ *  \left( {k_1 ,k_2 } \right) - \epsilon \left( {k_1 ,k_2 } \right)\epsilon \left( {k_1 ,k_2 } \right)} \right]} \right\}   \notag \\
 &= \Re \left\{ {\frac{1}{2}{\rm{E}}\left[ {\epsilon \left( {k_1 ,k_2 } \right)\epsilon ^ *  \left( {k_1 ,k_2 } \right)} \right]} \right\} = \sigma _{\epsilon _1 }^2 .
\end{align}
Q.E.D.

\section{Proof of Proposition 2}
Clearly 
\begin{align}
 Y &= X^2 + \beta X + \beta ^2/4 - \beta ^2/4  \notag  \\
    &= \left( X + \beta /2 \right)^2 - \beta ^2/4 \ge - \beta ^2/4 .
\end{align} 
Hence, the CDF of $Y$ is $F_Y(y) = 0$, if $y < -\beta ^2/4$. Then, if $y \ge -\beta ^2/4$, the CDF of $Y$ is as in \eqref{eq52}.
Since $X\rightarrow \mathcal{N}\left( 0,1 \right)$, the CDF is obtained as in \eqref{eq53}.
Finally, we conclude that the CDF of $Y$ is given as in \eqref{eq54}.
\setcounter{TempEqCnt}{\value{equation}}
\setcounter{equation}{51}
\begin{figure*}[ht]
  \normalsize
  \vspace*{-12pt}
\begin{align} \label{eq52}
F_Y\left( y \right) &=\text{Pr}\left( Y\le y \right) =\text{Pr}\left( X^2+\beta X\le y \right) =\text{Pr}\left( \left( X+\beta /2 \right) ^2\le y+\left( \beta /2 \right) ^2 \right)   \notag \\
&=\text{Pr}\left( -\sqrt{y+\left( \beta /2 \right) ^2}-\beta /2\le X\le \sqrt{y+\left( \beta /2 \right) ^2}-\beta /2 \right) .
\end{align}
\begin{align} \label{eq53}
F_Y\left( y \right) &=\left( 1 - \frac{1}{2}\text{erfc}\left( \frac{\sqrt{y+\left( \beta /2 \right) ^2}-\beta /2}{\sqrt{2}} \right) \right) -\left( 1- \frac{1}{2}\text{erfc}\left( \frac{-\sqrt{y+\left( \beta /2 \right) ^2}-\beta /2}{\sqrt{2}} \right) \right)  \notag \\
&=\frac{1}{2}\left[ \text{erfc}\left( \frac{-\sqrt{y+\left( \beta /2 \right) ^2}-\beta /2}{\sqrt{2}} \right) - \text{erfc}\left( \frac{\sqrt{y+\left( \beta /2 \right) ^2}-\beta /2}{\sqrt{2}} \right) \right] .
\end{align} 
\begin{equation}   \label{eq54}
F_Y \left( y \right) = \left\{ {\begin{array}{*{20}c}
   {\frac{1}{2}\left[ {\rm{erfc}}\left( {\frac{{ - \sqrt {y + \left( {\beta /2} \right)^2 }  - \beta /2}}{{\sqrt 2 }}} \right) - {{\rm{erfc}}\left( {\frac{{\sqrt {y + \left( {\beta /2} \right)^2 }  - \beta /2}}{{\sqrt 2 }}} \right)  } \right],} & {y \ge  - \frac{{\beta ^2 }}{4}}  \\
   {0,} & {y <  - \frac{{\beta ^2 }}{4}}  \\
\end{array}} \right. .
\end{equation}
    \hrulefill
\end{figure*}
Q.E.D.

\section{Derivation of Test Statistics for Decision Tree}
At the top-level node, the vector ${\bf{v}}_i$ is given as
\begin{align}
{\bf{v}}_i^{C_1} =& \frac{1}{{\sqrt {N'/{\rm{8}}} }}\sum\limits_{j = iN'/{\rm{8}} + 1}^{(i + 1)N'/{\rm{8}}} {{\bf{r}}\left( {{\rm{8}}j - {\rm{7}},{\rm{8}}j - {\rm{3}}} \right)} \notag \\
& {\rm{ + }}\frac{1}{{\sqrt {N'/{\rm{8}}} }}\sum\limits_{j = iN'/{\rm{8}} + 1}^{(i + 1)N'/{\rm{8}}} {{\bf{r}}\left( {8j - 6,8j - 2} \right)}  \notag  \\
& {\rm{ + }}\frac{1}{{\sqrt {N'/{\rm{8}}} }}\sum\limits_{j = iN'/{\rm{8}} + 1}^{(i + 1)N'/{\rm{8}}} {{\bf{r}}\left( {8j - 5,8j - 1} \right)}  \notag \\
& {\rm{ + }}\frac{1}{{\sqrt {N'/{\rm{8}}} }}\sum\limits_{j = iN'/{\rm{8}} + 1}^{(i + 1)N'/{\rm{8}}} {{\bf{r}}\left( {8j - 4,8j} \right)}   
\end{align}
and the estimated covariance matrix of the error is rewritten as
\begin{equation}
{{{\bf{\hat \Psi }}}^{C_1}} = \frac{1}{{N - 10}}\sum\limits_{k = 1}^{N - 9} {{\bf{I}}_{2D} \cdot \left[ {{\bf{r}}\left( {k,k + 9} \right) \circ {\bf{r}}\left( {k,k + 9} \right)} \right]} .
\end{equation}
Then, the test statistic is constructed as follows
\begin{equation}
{{\cal U}^{{C_1}}} = \sum\limits_{i = 0}^{G - 1} {{{\left( {{\bf{v}}_i^{{C_1}}} \right)}^T}{{\left( {{\bf{\hat \Psi }}{^{{C_1}}}} \right)}^{ - 1}}{\bf{v}}_i^{{C_1}}}. 
\end{equation}

At the middle-level node, the vector ${\bf{v}}_i$ is given by
\begin{align}
{\bf{v}}_i^{C_2} =& \frac{1}{{\sqrt {N'/{\rm{4}}} }}\sum\limits_{j = iN'/{\rm{4}} + 1}^{(i + 1)N'/{\rm{4}}} {{\bf{r}}\left( {{\rm{4}}j - {\rm{3}},{\rm{4}}j - {\rm{1}}} \right)} \notag \\
 &{\rm{ + }}\frac{1}{{\sqrt {N'/{\rm{4}}} }}\sum\limits_{j = iN'/{\rm{4}} + 1}^{(i + 1)N'/{\rm{4}}} {{\bf{r}}\left( {4j - 2,4j} \right)} 
\end{align}
and the estimated covariance matrix of the error is as follows
\begin{equation}
{{{\bf{\hat \Psi }}}^{C_2}} = \frac{1}{{N - 6}}\sum\limits_{k = 1}^{N - 5} {{\bf{I}}_{2D} \cdot \left[ {{\bf{r}}\left( {k,k + 5} \right) \circ {\bf{r}}\left( {k,k + 5} \right)} \right]} .
\end{equation}
The test statistic is constructed as follows
\begin{equation}
{{\cal U}^{{C_2}}} = \sum\limits_{i = 0}^{G - 1} {{{\left( {{\bf{v}}_i^{{C_2}}} \right)}^T}{{\left( {{\bf{\hat \Psi }}{^{{C_2}}}} \right)}^{ - 1}}{\bf{v}}_i^{{C_2}}}. 
\end{equation}

At the bottom-level node, the vector ${\bf{v}}_i$, the estimated covariance matrix of the error, and the test statistic have been defined in \eqref{eq18}, \eqref{eq19} and \eqref{eq20} respectively.

\section{Proof of Proposition 3}

Suppose that the mean of $\left|{\boldsymbol{\epsilon }}(k_1,k_2) \right|$ is ${\boldsymbol{\mu }} = {\rm{E}}\left[ {\left| {\boldsymbol{\epsilon }}(k_1,k_2) \right|} \right] = \left[ {\mu _1 , \cdots ,\mu _{2D } } \right]^T$, and the covariance matrices ${\bf{\Psi }}$ and ${\bf{\Phi }}$ are ${\bf{\Psi }} = {\rm{diag}}\left( {\sigma _{\epsilon _1 }^2, \cdots ,\sigma _{\epsilon _{2D} }^2 } \right)$ and ${\bf{\Phi }} = {\rm{diag}}\left( {\sigma _{\left|\epsilon \right|_1 }^2, \cdots ,\sigma _{\left|\epsilon \right| _{2D} }^2 } \right)$, respectively. From the \emph{Proof of Proposition 1},
\begin{align}
\sigma _{\left| \epsilon  \right|_1 }^2 &= {\rm{  E}}\left[ {\left( {\Re \left\{ {\epsilon \left( {k_1 ,k_2 } \right)} \right\}} \right)^2 } \right] - \left( {{\rm{E}}\left[ {\left| {\Re \left\{ {\epsilon \left( {k_1 ,k_2 } \right)} \right\}} \right|} \right]} \right)^2  \notag \\
&= \sigma _{\epsilon _1 }^2  - \mu _1^2. 
\end{align}
Clearly, we have $\mu _1^2  \ge 0$, $\sigma _{\epsilon _1 }^2  \ge 0$, $\sigma _{\left| \epsilon  \right|_1 }^2  \ge 0$ and it follows that $\sigma _{\epsilon _1 }^2  \ge \mu _1^2 $, $\sigma _{\epsilon _1 }^2  \ge \sigma _{\left| \epsilon  \right|_1 }^2 $. Similarly, all other elements in ${\bf{\Psi }}$ are greater than or equal to the corresponding elements in ${\bf{\Phi }}$. Therefore, we have ${\bf{\Psi }} \ge {\bf{\Phi }}$.
Q.E.D.

%\section{}

%\section*{Acknowledgment}

\ifCLASSOPTIONcaptionsoff
  \newpage
\fi

\bibliographystyle{IEEEtran}

\bibliography{./gao}

\end{document}